\documentclass[usenatbib]{mn2e}
\usepackage{graphicx,times}
\usepackage{bm}
\usepackage{epsf}
\usepackage{bm}
\usepackage{amsmath}
\usepackage{amssymb}
\usepackage{color}
\usepackage[usenames,dvipsnames,svgnames,table]{xcolor}

\def\cred{\textcolor{black}}

\def\br{{\bf r}}
\def\n{\noindent}

\def\myC{{\cred{\cal C}}}
\def\be{\begin{equation}}
\def\ee{\end{equation}}
\def\ben{\begin{eqnarray}}
\def\een{\end{eqnarray}}
\def\beqa{\begin{eqnarray}}
\def\eeqa{\end{eqnarray}}

\def\oh{{\hat\Omega}}
\def\nn{{\nonumber}}
\def\la{\langle}
\def\ra{\rangle}

\newcommand{\calH}{{\cal H}}

\date{\today,~ $ $Revision: 0.9 $ $}

\begin{document}

\twocolumn


\title[Effects of Linear Redshift Space Distortions and Perturbation Theory on BAOs: A 3D Spherical Analysis]
{Effects of Linear Redshift Space Distortions and Perturbation Theory on BAOs: A 3D Spherical Analysis}

\author[Pratten \& Munshi]
{\parbox{\textwidth}{Geraint Pratten$^1$ and Dipak Munshi$^{1,2}$}	\vspace{0.5cm}\\
$^1$School of Physics and Astronomy, Cardiff University, Queen's, Buildings, 5 The Parade, Cardiff, CF24 3AA, UK \\
$^2$Department of Physics \& Astronomy, University of Sussex, Brighton, BN1 9QH, UK \\
}

\maketitle

\begin{abstract}
The Baryon Acoustic Oscillations (BAO) are features in the matter power spectrum on scales of order 100 - 150 $h^{-1} \textrm{Mpc}$ that promise to be a powerful 
tool to constrain and test cosmological models. The BAO have attracted such attention that future upcoming surveys have been designed with the BAO at the forefront 
of the primary science goals. Recent studies have advocated the use of a spherical-Fourier Bessel (sFB) expansion for future wide field surveys that cover both 
wide and deep regions of the sky necessitating the simultaneous treatment of the spherical sky geometry as well as the extended radial coverage. Ignoring the possible effects of growth, which is not expected to be significant at low redshifts, we present 
an extended analysis of the BAO's using the sFB formalism by taking into account the role of non-linearities 
and linear redshift distortions in the oscillations observed in the galaxy power spectrum.
The sFB power spectrum has both radial and tangential dependence and it has been shown that in the limit that we approach a deep survey the sFB power spectrum is 
purely radial and collapses to the Cartesian Fourier power spectrum. This radialisation of information is shown to hold even in the 
presence of redshift space distortions (RSD) and 
1-loop corrections to the galaxy power spectrum albeit with modified tangential and radial dependence.
As per previous studies we find that the introduction of non-linearities leads to a damping of the oscillations in the matter power spectrum.
\end{abstract}


\section{Introduction}
Observations of the cosmic microwave background (CMB) and large-scale structure (LSS) will carry complementary cosmological information. While all-sky CMB observations, 
such as NASA's WMAP\footnote{ http://map.gsfc.nasa.gov/ } or ESA's Planck\footnote{ http://sci.esa.int/planck } experiments, primarily probe the distribution of matter and radiation at redshift $z = 1300$, large scale surveys such as ESA's 
Euclid\footnote{http://sci.esa.int/euclid} or the Square Kilometer Array (SKA)\footnote{http://www.skatelescope.org/} will provide a window at lower redshifts on order $z \approx 0 - 2$. The study of large scale structure appears to be a promising candidate in the study of 
the influence and role of the dark sectors in the standard model of cosmology. One particular phenomena of interest are the Baryon Acoustic Oscillations (BAOs) 
that manifest themselves in the matter power spectrum of galaxy clusters on cosmological scales of order $100 h^{-1} \textrm{Mpc}$. 
These oscillations in the matter power spectrum are generated just before recombination through 
the interplay between a coupled photon-baryon fluid and gravitationally interacting dark matter \citep{Sunyaev70,Peebles70,Eisenstein05,Seo03,Seo07}.  

The scale of the peaks and oscillatory features of the BAOs promises to be an important cosmological tool that acts as a standard ruler 
from which we can investigate and constrain 
dark energy parameters (see \cite{Eisenstein05,Amendola05,Dolney06,Wang06} for a small selection or representative literature), neutrino masses \citep{Goobar06}, 
modified theories of gravitation \citep{Alam06,Lazkoz06} and deviations from the standard model of cosmology \citep{Garcia08,Garcia09,February12}). 
Significant attention has been devoted to the BAOs and they were first detected with SDSS \footnote{http://www.sdss.org/} data \citep{Eisenstein05,Adelman08} and in subsequent surveys 
\citep{Colless03,Percival07}. 

The BAOs have been studied using standard Fourier space decompositions \citep{Seo03,Seo07}, real space analysis 
\citep{Eisenstein05,Slosar09,Xu10,Juszkiewicz12}, in 2D spherical harmonics defined on thin spherical shells \citep{Dolney06}, 
but also in the sFB expansion \citep{Rassat12}. 
It is important to note that different frameworks will make use of different information 
and will therefore have different constraining power for different cosmological parameters emphasising the complementarity of mixed studies \citep{Rassat08}. 
Previous studies, having predominantly focused on projected 2D surveys, have discarded radial information by projecting galaxy positions 
into tomographic redshift bins however, such a loss of information could be avoided by adopting a full 3D description (e.g. \cite{Asorey12}). 

Upcoming large scale structure surveys will provide cover for both large and deep areas of the sky and this will necessitate a formalism that 
can provide a simultaneous treatment of both the spherical sky geometry as well as an extended radial coverage. 
A natural basis for such a survey is provided by the sFB
decomposition, (see \cite{Heavens95,Fisher95,Percival04,Castro05,Erdogdu06,Abramo10,Leistedt11,Shapiro11,Rassat12,Lanusse12,Asorey12} for an incomplete selection of 
literature on the subject). 
In this prescription we expand a 3D tracer field, such as the galaxy density contrast, using the radial $(k)$ and tangential (i.e. along the surface of a sphere) 
$(\ell)$ dependence. 

The galaxy matter power spectrum is conventionally modelled using cosmological perturbation theory (PT). The linear order results will be valid at large scales where non-linear growth of structure under gravitational instability can be neglected. At smaller scales it is no longer possible to neglect the non-linear growth of structure and we 
need to incorporate higher-order corrections to the matter power spectrum. There are a number of different approaches currently in the literature to tackle this problem and 
we will present a more detailed description later on. Non-linear galaxy clustering bias arises from a non-linear mapping between the underlying matter density field and 
observed collapsed objects (e.g. galaxies or dark matter haloes) and galaxy bias is, in essence, an isocurvature perturbation. 
Current literature has investigated more detailed prescriptions for galaxy bias such as the effects of primordial non-Gaussianity, scale dependence or non-local bias. 
Another form of non-linearity arises from RSD generated through the internal motion of galaxies within haloes. This effect is known as the Finger-of-God effect 
\citep{Jackson72} and is distinct from the linear RSD considered in this paper \citep{Kaiser87}.  It is also possible to investigate the role of 
non-Gaussian initial conditions, such as those generated in various inflationary models, and how this propagates non-linear corrections through to the growth 
of structure. The signatures of non-Gaussianity in these models will be distinctly different (e.g. a modified bispectrum) to the signatures of non-Gaussianity in 
models that have Gaussian initial conditions and are allowed to undergo gravitational collapse. 

Throughout this paper we will follow the construction outlined in \cite{Rassat12} and generalise the method to study the role of redshift space distortions (RSD) 
and the non-linear (NL) evolution of density perturbations. Previous investigations have used 
standard perturbation 
theory (SPT), galaxy bias models and Lagrangian perturbation theory (LPT) to characterise the role of various non-linear corrections to the BAO signal using the 
3D Fourier power
spectrum $P(k)$ \citep{Jeong06,Nishimichi07,Jeong09,Nomura08,Nomura09}. These nonlinear corrections can be reassessed within the sFB framework to aid our 
understanding of how real world effects can impact the radialisation of information. 

Recent 
work \citep{Asorey12} utilising the sFB formalism has focused on how to recover the full 3D clustering information 
including RSD from 2D tomography using the angular auto and cross spectra 
of different redshift bins. Traditionally, RSD measurements have been made through spectroscopic redshift 
surveys such as the 2dF Galaxy Redshift Survey \citep{Colless03} and the Sloan Digital Sky Survey \citep{York00} with 
photometric surveys often being neglected because of the loss of RSD through photometric redshift errors. Upcoming surveys, spectroscopic and photometric, such as 
the Dark Energy Survey (DES)\footnote{www.darkenergysurvey.org}, Euclid, SKA, 
Physics of the Accelerating Universe Survey (PAU)\footnote{www.pausurvey.org} \citep{Benitez09}, Large Synoptic Survey Telescope (LSST)\footnote{www.lsst.org} or 
the Panoramic Survey Telescope and Rapid Response System (PanStarrs)\footnote{pan-starrs.ifa.hawaii.edu} offer the 
possibility of investigating the BAO and RSD through angular or projected clustering measurements 
\cite{Benitez09,Nock10,Crocce10,Gaztanaga11,Laureijs11,Ross11}.

As RSD and distortions arising from an incorrect assumption for the underlying geometry are similar \citep{Alcock79} 
the analyses of RSD using 3D data has to be used in conjunction 
with geometrical constraints \citep{Samushia11}. As approaches based purely on angular correlation functions do not depend on the 
background cosmological model, the angular clustering measures will be considerably simpler. The sFB is something of a mid-point between these two 
approaches and will, in general, be sensitive to the choice of fiducial concordance cosmology. 
This paper is organised as follows. In \textsection\ref{sec:red} we discuss the sFB expansion.  
In \textsection\ref{sec:redshift} we outline the effect of linear RSD and \textsection\ref{sec:real}
is devoted to issues related to realistic surveys.  In \textsection\ref{sec:pert}
we consider perturbative corrections to linear real-space results and consider the structure of the sFB spectra.
Results are discussed in 
\textsection\ref{sec:res} and conclusions presented in  
\textsection\ref{sec:conclu}. Discussions about finite size of the survey and discrete
sFB transforms are detailed in the appendices.

Throughout we  will adopt the WMAP 7 cosmological parameters \citep{WMAP7}: $h = 0.7, \Omega_b h^2 = 0.0226, \Omega_c h^2 = 
0.112, \Omega_{\Lambda} = 0.725, \sigma_8 = 0.816$. 

\section{Spherical Fourier-Bessel (sFB) Expansion}
\label{sec:red}
\subsection{Theory}
Spherical coordinates are a natural choice for the analysis of cosmological data as they can, by an appropriate choice of basis, be used to place an observer 
at the origin of the analysis. Upcoming wide-field BAO surveys will provide both large and deep coverage of the sky and we therefore require a simultaneous treatment 
of the extended radial coverage and spherical sky geometry. For this problem, the sFB expansion is a natural basis for the analysis of random fields 
in such a survey. 

We introduce a homogeneous 3D random field $\Psi (\hat{\Omega}, r)$ with $\hat{\Omega}$ defining a position on the surface of a sphere and $r$ denoting 
the comoving radial distance. The eigenfunctions of the Laplacian operators are constructed from products of the spherical Bessel functions of the first kind 
$j_{\ell} (kr)$ and spherical harmonics $Y_{\ell m}(\hat{\Omega})$ with eigenvalues of $-k^2$ for a 2-sphere. 
Assuming a flat background Universe, the sFB decomposition of our random field \citep{Binney91,Fisher95,Heavens95,Castro05} 
is given by:

\begin{equation}
  \Psi (\hat{\Omega},r) = \sqrt{\frac{2}{\pi}} \int d k \displaystyle\sum_{\lbrace \ell m \rbrace} \Psi_{\ell m} (k) \, k \, j_{\ell} (kr) Y_{\ell m}(\hat{\Omega}) ,
\end{equation}
\n
and the corresponding inverse relation given by:
\begin{equation}
\Psi_{\ell m} (k) = \sqrt{\frac{2}{\pi}} \int d^3 {\bf{r}} \, \Psi ({\bf{r}}) \, k \, j_{\ell} (kr) Y^{\ast}_{\ell m} (\hat{\Omega}) .
\end{equation}

\n
In our notation, $\lbrace \ell m \rbrace$ are quantum numbers and k represents the wavenumber.\footnote{We follow the same conventions as 
\cite{Leistedt11, Rassat12,Castro05}
but have made the substitutions $f ({\bf{r}}) \rightarrow \Psi ( {\bf{r}} )$ and $W_{\ell} (k_1 , k_2) \rightarrow I^{(0)}_{\ell} (k_1 , k_2)$.} 

Note that the 3D harmonic coefficients, $\Psi_{\ell m} (k)$ are a function of the radial 
wavenumber $k$. This decomposition can be viewed as the spherical polar analogy to the conventional Cartesian Fourier decomposition defined by:
\begin{align}
\Psi ({\bf{r}}) &= \frac{1}{(2 \pi)^{3/2}} \int d^3 k \, \Psi (k) \, \rm{e}^{ i {\bf{k}} \cdot {\bf{r}}} , \\
\Psi ({\bf{k}}) &= \frac{1}{(2 \pi)^{3/2}} \int d^3 x \, \Psi ({\bf{r}}) \, \rm{e}^{ - i {\bf{k}} \cdot {\bf{r}}} .
\end{align}
The Fourier power spectrum, $P_{\Psi \Psi}$, is defined as the 2-point correlation function of the Fourier coefficients $\Psi (k)$:
\begin{equation}
  \langle \Psi ({\bf{k}}) \Psi^{\ast} ({\bf{k}}^{\prime}) \rangle = (2 \pi)^3 P_{\Psi \Psi} (k) \delta^3 \left( {\bf{k}} - {\bf{k}}^{\prime} \right) . 
\end{equation}
Similarly we can define a 3D sFB power spectrum, $C_{\ell} (k)$, of our random field by calculating the 2-point correlation function of the 3D harmonic coefficients:
\begin{equation}
  \langle \Psi_{\ell m} (k) \Psi^{\ast}_{\ell^{\prime} m^{\prime}} (k^{\prime}) \rangle = \myC_{\ell} (k) \delta^{1\rm D}(k-k^{\prime}) 
  \delta^{\rm K}_{\ell \ell^{\prime}} \delta^{\rm K}_{m m^{\prime}} . 
\end{equation}
It is possible to relate the Fourier coefficients $\Psi ({\bf{k}})$ with their sFB analog $\Psi_{\ell m} (k)$ through the following expression
\begin{equation}
  \Psi_{\ell m}(k) = \frac{i^{\ell} k}{( 2 \pi )^{3/2}} \int d \Omega_{k} \Psi ({\bf{k}}) Y_{\ell m} ( \hat{\Omega}_k )
\end{equation}
where the angular position of the wave vector ${\bf{k}}$ in Fourier space is denoted by the unit vector $\hat{\Omega} (\theta_k , \phi_k )$. 
\n
The Rayleigh-expansion of a plane wave is particularly useful in connecting the spherical harmonic description with the 3D Cartesian expression. The second 
expression we present here is derived by differentiating the first and will be used in the derivation of RSD: 
\ben
&& {\rm{e}}^{i {\bf{k}} \cdot {\bf{r}}} = 4 \pi \displaystyle\sum_{\ell m} i^{\ell} \, j_{\ell} (kr) Y_{\ell m} (\hat{\Omega}_k) Y_{\ell m} (\hat{\Omega}); \\
&& i ( \hat{\Omega}_k \cdot \hat{\Omega} ) {\rm{e}}^{i {\bf{k}} \cdot {\bf{r}}} = 4 \pi 
\displaystyle\sum_{\ell m} i^{\ell} \, j^{\prime}_{\ell} (kr) Y_{\ell m} (\hat{\Omega}_k) Y_{\ell m} (\hat{\Omega}) .
\een
In general the radial eigenfunctions are ultra-spherical Bessel functions but they can be approximated by spherical Bessel functions when the curvature of the Universe is small (e.g. \cite{Zaldarriaga98}). Throughout this paper we will use $j^{\prime}_{\ell} (x)$ and $j^{\prime \prime}_{\ell} (x)$ to denote the first 
and second derivatives of $j_{\ell} (x)$ with respect 
to its argument $x$. 
The expressions for the first and second derivatives are given in Eq.(B2) and Eq.(B3). Imposing a finite boundary condition on the radial direction 
will result in a discreet sampling of the k-modes. This will be discussed in more detail later. 

\subsection{Finite Surveys}
In order to consider realistic cosmological random fields, such as the galaxy density contrast, we need to take into account the partial observation effects 
arising from finite survey volumes. Concise discussions of this point are given in \citep{Rassat12,Asorey12} and as such we 
will not devote much time to this point referring the reader to the given references.

The selection function simply denotes the probability of including a galaxy within a given survey. An observed random 
field $\Psi^{\textrm{obs}}({\bf{r}})$ can be related to an underlying 3D random field through a survey-dependent radial selection function $\phi(r)$ that modulates the 
underlying field:

\begin{equation}
  \Psi^{\textrm{obs}} ({\bf{r}}) = \phi(r) \Psi ({\bf{r}}) .
\end{equation}

\n
It is possible to introduce an analogous tangential selection function but we will, as per \cite{Rassat12}, neglect this possibility assuming that we have full sky coverage. The resulting sFB power spectrum is given by

\begin{equation}
{\mathcal{C}}^{(00),\textrm{obs}}_{\ell} (k_1 , k_2) = \left( \frac{2}{\pi}\right )^2 \int k^{\prime 2} d k^{\prime} \, 
    I^{(0)}_{\ell} (k_1,k^{\prime}) I^{(0)}_{\ell} (k_2,k^{\prime}) P_{\delta \delta}(k^{\prime}) 
 \label{eqn:AngPowSpec}
\end{equation}
where the modified window function is given by:
\begin{equation}
\label{eq:Window}
I^{(0)}_{\ell} (k,k^{\prime}) = \int dr \, r^2 \phi(r) \, k \, j_{\ell} (kr) \, j_{\ell} (k^{\prime} r) . 
\end{equation}
The sFB power spectrum tends to rapidly decay as we move away from the diagonal $k=k^{\prime}$ and it will often be much more useful to focus purely on the diagonal 
contribution $C^{(00)}_{\ell} (k,k)$.

\section{Redshift Space Distortions}
\label{sec:redshift}
The measured distribution of galaxies is not without limits though as various systematic and survey dependent errors become more important. 
In practice, the observed galaxy redshift distributions are distorted due to the peculiar velocity of each galaxy. The anisotropies generated 
by the peculiar velocities are known as \emph{redshift space distortions}. Although this distortion of the measured redshifts will necessarily 
complicate the cosmological interpretation of the spectroscopic galaxy surveys, RSD are currently one of the most 
optimistic probes for the measurement of the 
growth rate of structure formation and, as a result, an interesting probe of models for dark energy and modified theories of gravity. 

The effect of RSD on the matter power spectrum can be split into two effects, the Kaiser effect and the FoG effect. 
The Kaiser effect 
corresponds to the coherent distortion of the peculiar velocity along the line of sight with an amplitude controlled by the growth-rate parameter, 
leading to an enhancement of the power spectrum amplitude at small k \citep{Kaiser87}. 
The FoG
effect arises due to the random distribution of peculiar velocities leading to an incoherent contribution in which dephasing occurs and the clustering 
amplitude is suppressed \citep{Jackson72}. It is thought that the suppression of the amplitude is particularly important around the size of halo forming regions, i.e. at 
large k \citep{Taruya10}. 

For an isotropic structure in linear theory, the Kaiser effect means that an observer will measure more power in the radial direction 
than in the transverse modes. The amplitude of this distortion is modulated by the distortion parameter

\begin{equation}
 \beta = \frac{f (\Omega_0)}{b(z)} = \frac{1}{b(z)} \frac{d \ln D (a)}{d \ln a} \approx \frac{\Omega^{\gamma}_m  (a)}{b(z)} 
\end{equation}

\n
where:

\begin{equation}
 \Omega_m (a) = \frac{\Omega_{m,0}}{a^3} \frac{H^2_0}{H^2 (a)} 
\end{equation}

\n
such that $a$ is the scale factor, $H(a)$ is the Hubble parameter, $H_0$ is the Hubble parameter at present time and $D(z)$ the linear growth factor for which $f (z) \equiv d \ln D / d \ln a$.
In this parameterisation, $\gamma$ is directly related to our theory of gravitation such that General Relativity predicts $\gamma \simeq 0.55$ and $\Omega_m$ is 
the usual mass density parameter \citep{Wang98,Linder05}. 
This means that RSD can be used to probe the growth of structure, the galaxy clustering bias function $b(z)$ as well as probing dark energy and modified theories 
of gravity \citep{Guzzo08}. Measuring the growth rate from RSD is a non-trivial procedure and a detailed understanding of systematic errors is 
crucial in order to disentangle different theories of gravity or dark energy \cite{Torre12}. Euclid aims to constrain the growth 
rate parameter to the percent level 
but incomplete modelling of RSD introduces systematics on order $10-15 \%$ \citep{Taruya10,Okumura11,Bianchi12,Torre12}. This makes 
the study of RSD in the sFB formalism all the more timely. In this next section we will outline some of the basic ingredients that are used in modelling RSD 
in Fourier space before constructing the analogous results in the sFB formalism. 

\subsection{RSD in Fourier Space}
Before presenting the RSD in the sFB formalism we briefly review some of the key results from modelling RSD in Fourier space and 
the appropriate limitations that are adopted in the model. 

The effect of a peculiar velocity ${\bf{v}}$ is to distort the apparent comoving position ${\bf{s}}$ of a galaxy from its true comoving position ${\bf{r}}$:

\begin{align}
 {\bf{s}} &= {\bf{r}} + \frac{ v_{\parallel} ( {\bf{r}} ) \hat{n}}{a H (a) } \nonumber \\
 &= {\bf{r}} + f \phi ( {\bf{r}} ) \hat{n} 
\end{align}

\n
where $f$ is the linear growth rate, $\hat{n}$ is a vector lying parallel to an observer's line of sight 
and $v_{\parallel}$ is the component of the velocity parallel to the line of sight. The 
resulting redshift space density field $\delta_s ({\bf{s}})$ is obtained by imposing mass conservation, $\left[ 1 + \delta_s ({\bf{s}}) \right] d^3 {\bf{s}} 
= \left[ 1 + \delta_r ) {\bf{r}} ) \right] d^3 {\bf{r}}$, which results in the following:

\begin{equation}
 \left[ 1 + \delta_s ({\bf{s}}) \right] = \left[ 1 + \delta_r ({\bf{r}}) \right] \left| \frac{ d^3 {\bf{s}}}{d^3 {\bf{r}}} \right|^{-1} .
\end{equation}

\n
To simplify the analysis we can adopt the distant observer approximation in which we neglect the curvature of the sky and the Jacobian reduces to a 
term relating only to the line of sight

\begin{equation}
 \frac{ \partial s }{ \partial r} = 1 + f \phi^{\prime}
\end{equation}

\n
where a prime denotes differentiation with respect to the line of sight, i.e. parallel to $\hat{n}$:

\begin{equation}
 \phi^{\prime} ( {\bf{r}} ) = \partial_{\parallel} \left[ \frac{v_{\parallel}}{ f a H(a)} \right] . 
\end{equation}

\n
The redshift space density contrast can be re-written as:

\begin{equation}
 \delta_s ( {\bf{s}} ) = \frac{ ( \delta ({\bf{r}}) - f \phi^{\prime} ({\bf{r}}) )}{( 1 + f \phi^{\prime} {\bf{r}} )} .
\end{equation}

\n
Assuming an irrotational velocity field with a velocity divergence field $\theta ({\bf{r}}) = \nabla \cdot {\bf{v}} ({\bf{r}})$ we obtain the following useful 
relationship, $\phi ({\bf{r}}) = - ( \nabla^{-1} \theta ({\bf{r}}) )^{\prime}$. In Fourier space these equations simplify as 
$\phi^{\prime} (k) = - \mu^2 \theta (k)$, where we have made use of the fact that $( \nabla^{-1} )^{\prime \prime} 
= ( k_{\parallel} / k )^2 = \mu^2$. In our notation $k_{\parallel}$ denotes the modes parallel to the line of sight and $k_{\perp}$ denotes modes perpendicular 
to the line of sight where $k^2 = k^2_{\parallel} + k^2_{\perp}$.
\cite{Scoccimarro99} the redshift space density field can be written as 

\begin{align}
 \delta_s ( k , \mu ) &= \int \frac{ d^3 {\bf{s}}}{( 2 \pi)^3} e^{- i {\bf{k}} \cdot {\bf{s}}} \delta_s ( {\bf{s}} ) \nonumber \\
 &= \int \frac{d^3 {\bf{r}}}{(2 \pi)^3} e^{- i {\bf{k}} \cdot {\bf{r}}} e^{- i k f \mu} \left[ \delta( {\bf{r}} ) + f \mu^2 \theta ({\bf{r}}) \right]
\end{align}

\n
and the corresponding power spectrum as:

\begin{align}
P_s (k,\mu) &= \int \frac{ d^3 {\bf{r}}}{(2 \pi)^3} e^{- {\bf{k}} \cdot {\bf{r}}} \Big\langle e^{- i k f \mu ( \phi({\bf{r}}) - \phi({\bf{r}}^{\prime})) } 
\nonumber \\ &\times 
\left[ \delta ({\bf{r}}) + f \mu^2 \theta ({\bf{r}}) \right] \left[ \delta ({\bf{r}}^{\prime}) + f \mu^2 \theta ( {\bf{r}}^{\prime} ) \right] \Big\rangle .
\end{align}

\n
This prescription for the Fourier power spectrum has been constructed in the plane-parallel or distant observer approximation. The terms in the square brackets 
is the conventional Kaiser effect as described earlier. The exponential prefactor corresponds to the small-scale velocity dispersion and relates to the 
Fingers-of-God effect described earlier. A simplified phenomenological power spectrum was derived by \cite{Scoccimarro04} by assuming that the exponential 
prefactor may be separated from the ensemble average

\begin{equation}
 P_s (k , \mu ) = e^{-(f k \mu \sigma_v )^2} \left[ P_{\delta \delta} (k) + 2 f \mu^2 P_{\delta \theta} (k) + f^2 \mu^4 P_{\theta \theta} (k) \right] ,
\end{equation}

\n
where $\sigma_v$ is a velocity dispersion defined in \cite{Scoccimarro04}. In the linear regime we have $P_{\delta \delta} = P_{\delta \theta} = P_{\theta 
\theta}$ and the velocity dispersion prefactor tends towards zero. In such a limit we simply recover the linear result of \cite{Kaiser87}:

\begin{equation}
 P_s (k,\mu) = \left[ 1 + 2 f \mu^2 + f^2 \mu^4 \right] P_{\delta \delta} (k) .
\end{equation}

\n
Such a limit corresponds to making a number of approximations. For example, we require that the velocity gradient is sufficient small, the density and 
velocity perturbations must be accurately described by the linear continuity equations, the real-space density perturbations are well described by the linear 
results, i.e. $\delta ({\bf{r}}) \ll 1$, such that higher-order contributions are suppressed and we also require that the small-scale velocity dispersion 
tends towards zero and may be neglected. Such approximations appear to hold on the largest scales and a lot of distortion features are well modelled by 
this approximation. It is however known that this theory breaks down as we approach the quasi-linear and non-linear regimes. The result of \cite{Scoccimarro04} 
makes certain approximations about the separability of the exponential prefactor which neglects possible coupling terms between the velocity and density 
fields. A lot of effort has been invested in constructing non-linear models for RSD and upcoming surveys should prove to be a 
fruitful testing ground for many of these models \cite{Hivon95,Scoccimarro99,Scoccimarro04,Crocce08,Matsubara08a,Matsubara08b,
Taruya09,Taruya10,Matsubara11,Okamura11b,Sato11,Torre12}. 
We construct the RSD in the sFB formalism by first working to the linear Kaiser result and exploring the 
phenomenology of such an extension. 

\subsection{RSD in sFB Space}
As previously mentioned, the effect a peculiar velocity, or a departure from the Hubble flow, $v({\bf{r}})$ at ${\bf{r}}$ is to introduce a distortion to the galaxy positions in the 
redshift space ${\bf{s}}$:
\begin{equation}
  {\bf{s}}({\bf{r}}) = {\bf{r}} + {\bf{v}}({\bf{r}}) \cdot \hat{\Omega} . 
\end{equation}
We denote the harmonics of a field $\Psi ({\bf{r}})$ when convolved with a selection function, $\phi(s)$, by $\tilde{\Psi}_{lm} (k)$. These harmonics take into 
account the RSD:
\begin{equation}
 \tilde{\Psi}_{\ell m}(k) = \sqrt{\frac{2}{\pi}} \int s^2 ds \, \int d\hat{\Omega} \, \phi(s) \Psi({\bf{r}}) \, k \, j_{\ell} (ks) Y^{\ast}_{\ell m}(\hat{\Omega}) .
\end{equation}
The Fourier transform of the linearised Euler equation can be used to relate the Fourier transform of the density contrast, $\delta ({\bf{k}})$, to that of the 
peculiar velocity field $v({\bf{r}})$:
\begin{equation}
  {\bf{v}}({\bf{k}}) = - i \beta {\bf{k}} \frac{\delta ({\bf{k}})}{k^2} 
\end{equation}
where $b$ is the linear bias parameter. Following the procedure outlined in \cite{Heavens95}, we can establish a series expansion in $\beta$ such that the lowest 
order coefficients $\Psi^{(0)}_{\ell m} (k)$ are obtained by neglecting the RSD:
\begin{align}
  &\tilde{\Psi}_{\ell m} (k) = \tilde{\Psi}^{(0)}_{\ell m} (k) + \tilde{\Psi}^{(1)}_{\ell m} (k) + \dots ; \\
  &\tilde{\Psi}^{(0)}_{\ell m} (k) = \sqrt{\frac{2}{\pi}} \int^{\infty}_0 k^{\prime} dk^{\prime} \, \Psi_{\ell m} (k^{\prime}) I^{(0)}_{\ell} 
  (k^{\prime} , k ); \label{eq:PsiO0} \\
  &\tilde{\Psi}^{(1)}_{\ell m} (k) = \sqrt{\frac{2}{\pi}} \int^{\infty}_0 k^{\prime} dk^{\prime} \, \Psi_{\ell m} (k^{\prime}) I^{(1)}_{\ell} 
  (k^{\prime} , k ) \label{eq:PsiO1}.
\end{align}
The kernels $I^{(0)}_{\ell} (k^{\prime} , k )$ and $I^{(0)}_l (k^{\prime} , k )$ define the convolution and are dependent on the choice of selection function. 
Note that $I^{(0)}_{\ell} (k^{\prime} , k )$ is simply the window function we encountered previously in Eq.(\ref{eq:Window}). The kernels can be shown to be:
\begin{align}
 &I^{(0)}_{\ell} (k,k^{\prime}) = \int dr \, r^2 \phi(r) \, k \, j_{\ell} (kr) \, j_{\ell} (k^{\prime} r)  \\
 &I^{(1)}_{\ell} (k,k^{\prime}) = \frac{\beta}{k^{\prime}} \int dr \, r^2 \, k \, \frac{d}{dr} \left( \phi(r) j_{\ell} (kr) \right) \, 
 j^{\prime}_{\ell} (k^{\prime} r) . 
  \label{eqn:Il1}
\end{align}
\n
The lowest order corrections due to RSD are therefore encapsulated in $\Psi^{(1)}_{lm} (k)$. We can define a set of power spectra by 
using these harmonic coefficients:
\begin{align}
  \langle \Psi^{\alpha}_{\ell m} (k) \Psi^{\beta \ast}_{{\ell}^{\prime}m^{\prime}} (k^{\prime}) \rangle &= \myC^{(\alpha \beta)}_{\ell} 
  (k,k^{\prime}) \delta_{1D}(k-k^{\prime}) 
      \delta_{\ell {\ell}^{\prime}} \delta_{m m^{\prime}} , \\
  \langle \Psi^{\alpha}_{\ell m} (k) \tilde{\Psi}^{\beta \ast}_{{\ell}^{\prime}m^{\prime}} (k^{\prime}) \rangle &= \tilde{\myC}^{(\alpha \beta)}_{\ell} 
  (k,k^{\prime}) 
      \delta_{\ell {\ell}^{\prime}} \delta_{m m^{\prime}} .
\end{align}
We can construct a generalised power spectrum by using the common structure between Eq.(\ref{eq:PsiO0}) and Eq.(\ref{eq:PsiO1}):
\begin{equation}
  \tilde{\myC}^{(\alpha \beta)}_{\ell} (k_1 , k_2) = \left( \frac{2}{\pi} \right)^2 \int k^{\prime 2} dk^{\prime} \, 
    I^{(\alpha)}_{\ell} (k_1,k^{\prime}) I^{(\beta)}_{\ell} (k_2,k^{\prime}) P_{\delta \delta} (k^{\prime}) . 
\label{eqn:PowSpec}
\end{equation}
\n
The total redshifted power spectrum will be given by a sum of the various contributions:
\begin{equation}
  \tilde{\myC}_{\ell} (k_1,k_2) = \tilde{\myC}^{(00)}_{\ell} (k_1,k_2) + 2 \, \tilde{\myC}^{(01)}_{\ell} (k_1,k_2) + \tilde{\myC}^{(11)}_{\ell} (k_1,k_2) . 
\end{equation}
If we ignore the effects introduced by the selection function, i.e. set $\phi(r)$ = 1, then we recover the result for the unredshifted 
contributions \cite{Heavens95,Fisher95,Castro05}:
\begin{equation}
  \myC^{(00)}_{\ell} (k,k) = P_{\delta \delta} (k) . 
\end{equation}

\n
These expressions hold for surveys with all-sky coverage. In the presence of homogeneity and isotropy the 3D power spectrum will be independent of radial wave 
number $\ell$. The introduction of a sky mask breaks isotropy and introduces additional mode-mode couplings, the analysis will be generalised to this case in the next 
section. In the above equations we neglect a number of additional non-linear terms including General Relativistic corrections, velocity terms and 
lensing terms. It is also possible to adopt a full non-linear approach to RSD where the 
non-linear spectrum has significantly more complicated angular structure than in linear theory \cite{Shaw08}. The RSD information will be dependent on the relative 
clustering amplitude of the transverse modes and the radial modes, \cite{Asorey12}. Our ability to recover information and the extent to which the information 
radialises will naturally depend on the geometry of the survey and which modes we are able to include. 

\subsection{BAO Wiggles Only}
The BAOs can be 
isolated by constructing a ratio between the observed matter power spectrum $P^{\textrm{B}}_{\delta \delta} (k)$ and a theoretical matter power spectrum
$P^{\textrm{nB}}_{\delta \delta} (k)$ constructed from a zero-baryon (or no-wiggle) transfer function in which the oscillations do not show up \citep{Eisenstein05}. Using these 
two power spectra, the ratio $R^P (k)$ will reduce the dynamic range and isolates the oscillatory features of the BAOs:

\begin{equation}
 R^P (k) = \frac{P^B (k)}{P^{nB} (k)} .
\end{equation}

\n
This ratio is clearly defined for the Fourier space power spectrum but an appropriate generalisation to the sFB formalism may be 
constructed by calculating the ratio of the 
angular power spectra defined in Eq. (\ref{eqn:AngPowSpec}), with the matter power spectrum $C^B_{\ell} (k)$ to the angular power 
spectrum with the zero-Baryon power spectrum $C^{nB}_{\ell} (k)$ \citep{Rassat12}:

\begin{equation}
 R^C_{\ell} (k) = \frac{C^{B}_{\ell} (k)}{C^{nB}_{\ell} (k)} .
\end{equation}

\n
It is important to note that the characterisation method (i.e. how we choose to construct our ratio) can affect the characteristic scale of the BAOs when we 
take into account non-linear effects. This means that care has to be taken when comparing results that implement different methods \citep{Rassat08}. 
As an example we 
could construct our ratio by using the no-wiggles transfer function of \cite{Eisenstein05} or adopt an interpolation scheme to construct a smooth parametric curve 
\cite{Blake06,Percival07,Seo07}.
A different choice of smoothed matter power spectra, cosmological parameters, growth history or similar can impact the phenomenological behaviour of 
the underlying physics (e.g. location of BAO peaks). 
Other methods for characterising the acoustic oscillation scales can be found, for example, in \citep{Percival07b,Nishimichi07}. 

\subsection{Results: RSD}
In Figure [\ref{fig:RSD}] we compare $\tilde{C}_l (k)$ against a linear redshift space power spectrum, $P_s (k)$,
spectra for $\ell = 5,50$ at two given surveys corresponding to $r = 100 , 1400 h^{-1} \textrm{Mpc}$. In this plot the ratios are constructed by considering the differences between the appropriate spectra. The following ratios have been used:

\begin{align}
 R_{\ell}^{C,\textrm{RSD}} (k) &= \frac{C_{\ell}^{\textrm{RSD,Lin,B}} (k)}{C_{\ell}^{\textrm{RSD,Lin,nB}} (k)} \label{F1Ratio1} \\
 R_{\ell}^{C,\textrm{nRSD}} (k) &= \frac{C_{\ell}^{\textrm{nRSD,Lin,B}} (k)}{C_{\ell}^{\textrm{nRSD,Lin,nB}} (k)} \label{F1Ratio2} \\
 R^{P,\textrm{RSD}} (k) &= \frac{P^{\textrm{RSD,Lin,B}} (k)}{P^{\textrm{RSD,Lin,nB}} (k)} = R^{P,\textrm{nRSD}} (k)  \label{F1Ratio3}.
\end{align}

\begin{figure}
\begin{minipage}[b]{0.5\textwidth}
\begin{center}
\includegraphics[width=1\textwidth]{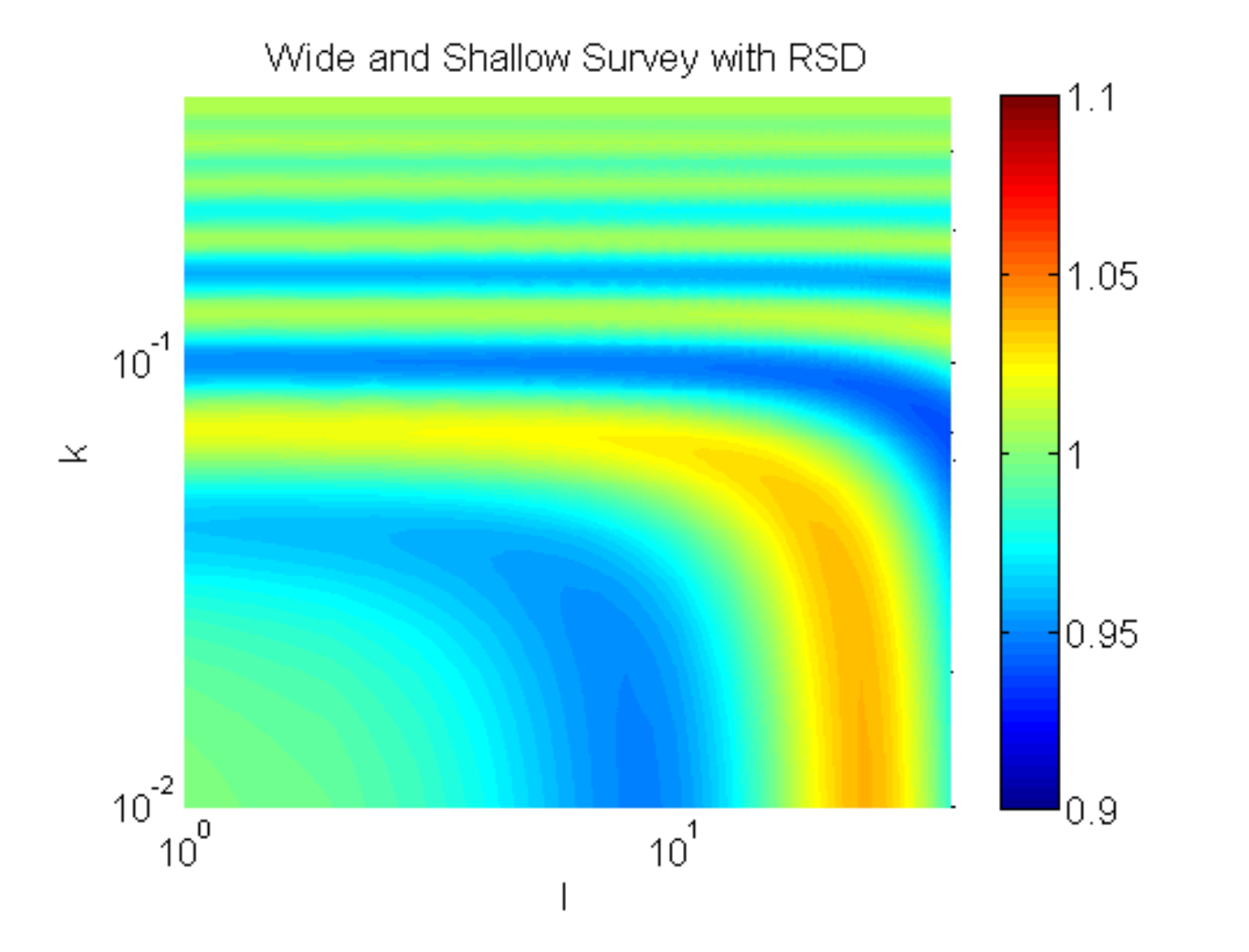}
\end{center}
\caption{Ratio $R^{C}_{\ell} (k)$ of sFB spectrum with and without the physical effects of baryons in $(\ell,k)$ phase space 
for a wide and shallow survey of $r_0 = 100 h^{-1} 
\textrm{Mpc}$ using a Gaussian selection function but with the inclusion of redshift space distortions. }
\label{fig:C-100-RSD}
\end{minipage}

\hspace{0.5cm}

\begin{minipage}[b]{0.5\textwidth}
\includegraphics[width=1\textwidth]{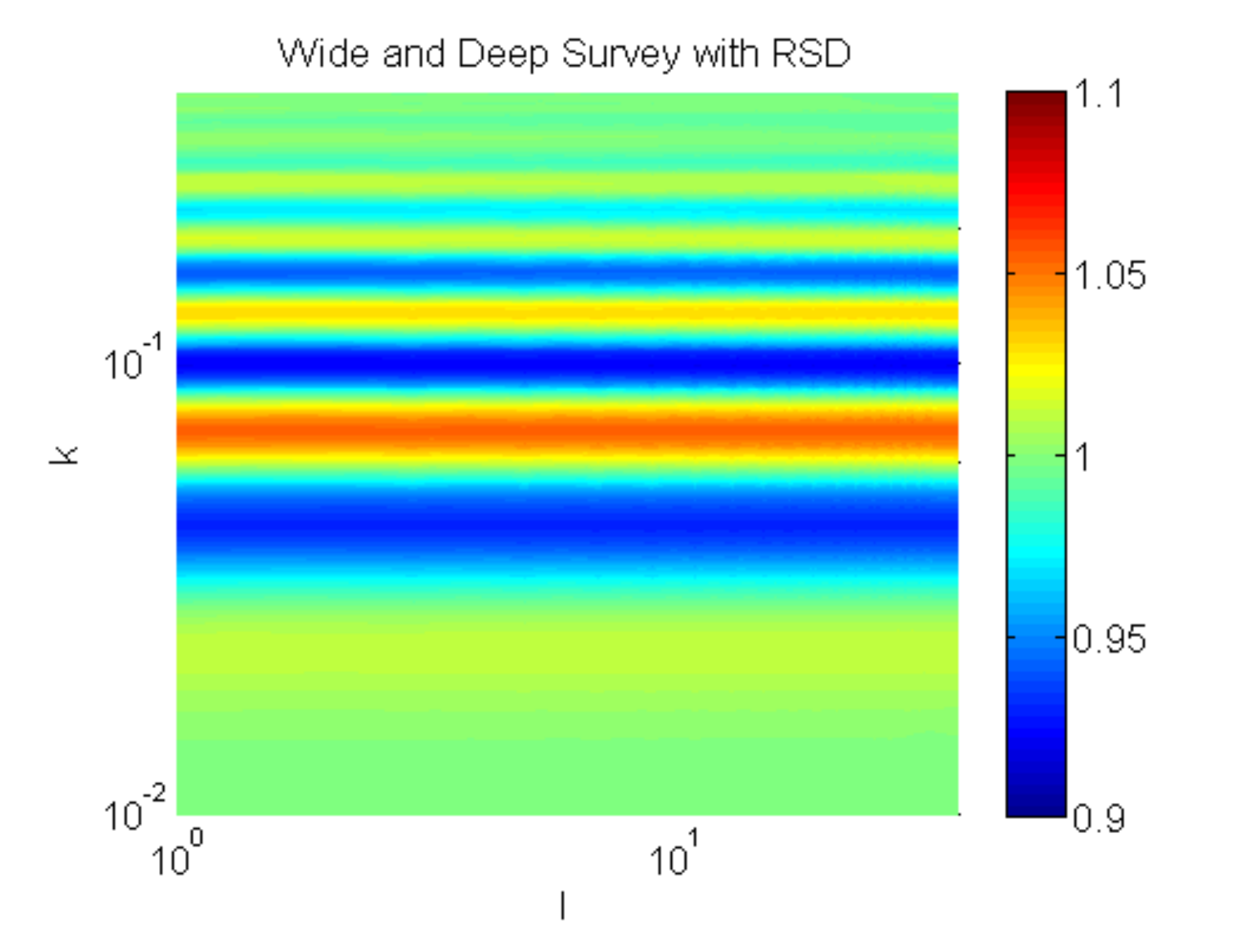}
\begin{center}
\end{center}
\caption{Ratio $R^{C}_{\ell} (k)$ of sFB spectrum with and without the physical effects of baryons in $(\ell,k)$ phase space for a 
wide and deep survey of $r_0 = 1400 h^{-1} 
\textrm{Mpc}$ using a Gaussian selection function but with the inclusion of redshift space distortions.}
\label{fig:C-1400-RSD}{}
\end{minipage} 
\end{figure}

\n
In Figure [\ref{fig:RSD}], the blue line corresponds to Eq.(\ref{F1Ratio1}), the purple line to Eq.(\ref{F1Ratio2}) and the red line to Eq.(\ref{F1Ratio3}). Figures [\ref{fig:C-100-RSD}, \ref{fig:C-1400-RSD}, \ref{fig:C-100-RSD-L-LDR}] correspond to Eq.(\ref{F1Ratio1}). 

The redshift space Fourier power spectrum is simply the result 
derived in \cite{Kaiser87} and corresponds to:

\begin{equation}
 P^s (k , \mu ) = \left[ 1 + 2 \mu^2 f + \mu^4 f^2 \right] P(k) .
\end{equation}

\n
In this linear limit, the redshift space ratio $R_s (k)$ tends to the real space ratio $R(k)$ as the linear prefactors corresponding to the redshift 
space corrections cancel. It is apparent that in Figure [\ref{fig:RSD}] the sFB spectra are damped relative to the power spectra. 
This arises due to mode-mixing contributions inherent when working with the sFB formalism. 
The unredshifted contributions are constructed from products of Bessel functions that form an orthogonal basis and there is no radial mode-mixing. When introducing RSD the higher-order terms are decomposed with respect to products involving derivatives of the spherical Bessel functions which does not form a perfectly orthogonal set of basis functions. As a result of RSD, off-diagonal elements will be generated and there is now coupling between modes. This radial mode-mixing is an intrinsic geometrical artifact of RSD on large scales and carries a distinctive damping signature \citep{Heavens95,Zaroubi96,Shapiro11}. Such a mode-mixing term is not present in the Kaiser analysis where the basis functions are plane waves which have well behaved derivatives that maintain the orthogonality of the basis. In the deep survey limit it 
is seen that the redshift space sFB spectra do tend towards their Fourier spectra counterparts in terms of the shape, amplitude and phase albeit with the presence of the distinctive damping generated by mode-mixing which is predominantly seen at small scales and hence large $k$. 

The effects of RSD can be seen in Figures [\ref{fig:C-100-RSD}-\ref{fig:C-1400-RSD}] in comparison to the equivalent configurations without the presence of RSD in Figures [\ref{fig:C-100-NRSD-L}-\ref{fig:C-1400-NRSD-L}]. A lower dynamical range comparison is presented in Figures [\ref{fig:C-100-NRSD-L-LDR}-\ref{fig:C-100-RSD-L-LDR}] to enhance the impact that RSD have on the BAO. Note the enhanced power at low $\ell$ and $k$ as well as some level of fuzziness introduced by the mode mixing. The peak amplitudes are damped at low $\ell$ and all the features can be seen in the corresponding slice plots of Figure [\ref{fig:RSD}]. In a future paper we will consider the hierarchy of multipole moments in Fourier space RSD and how measures constructed from the multipole moments can be related to RSD in the sFB formalism. 

\section{Realistic Surveys}
\label{sec:real}
The results that have been discussed above are somewhat idealised in the sense that we assume all-sky coverage with no noise. In realistic surveys we will often 
need to take into account the presence of a mask (relating to partial sky-coverage) and noise. If the noise is inhomogeneous we will be presented with a further 
complication. For partial sky coverage we find mode-mode couplings in the harmonic domain that result in the individual masked harmonics being described by 
a linear combination of our idealised all-sky harmonics. We do not discuss the role of partial sky-coverage in much detail but do present results generalising 
our formalism to include a survey mask. 

\subsection{Partial-Sky Coverage and Mode Mixing}
Large scale surveys do not, generally, have full-sky coverage. Instead the information regarding sky-coverage is encapsulated in a mask $\chi(\hat{\Omega})$ which 
is unity for areas covered in the survey and zero for regions outside the survey. The field harmonics are therefore modulated in the presence of a mask:
\begin{equation}
  \tilde{\Psi}_{\ell m} (k) = \sqrt{\frac{2}{\pi}} \int s^2 ds \int d\hat{\Omega} \, \left[ \phi (s) \chi (\hat{\Omega}) \right] \Psi({\bf{r}}) j_{\ell} (ks) 
  Y^{\ast}_{\ell m} 
      (\hat{\Omega}) . 
\end{equation}
The convolved power-spectra in the presence of the mask takes the following form:
\begin{align}
&\tilde{\myC}^{(\alpha \beta)}_{\ell} (k_1, k_2) = \left( \frac{2}{\pi} \right)^2 \displaystyle\sum_{\ell_a} \displaystyle\sum_{\ell_b} \int k^{\prime} dk^{\prime} 
      \int k^{\prime \prime} dk^{\prime \prime} \nn \\
 &\qquad \times W^{(\alpha)}_{\ell \ell_a} (k_1, k^{\prime}) W^{(\beta)}_{\ell \ell_b} (k_2, k^{\prime \prime}) \frac{I_{\ell \ell_a \ell_b}}{(2\ell+1)} 
 \myC^{\chi}_{\ell_a} \myC_{\ell_b} (k^{\prime},
      k^{\prime \prime}) ; \\
&\quad\quad\quad\myC^{\chi}_{\ell} = \langle \chi_{\ell m} \chi^{\ast}_{\ell m} \rangle 
\end{align}
where $I_{\ell_1 \ell_2 \ell_3}$ is the Gaunt integral (see Eq.(\ref{eq:Gaunt}) of Appendix-B). The convolved power spectrum is a linear combination of all-sky spectra and depends on the power spectrum 
of the adopted mask (see Appendix-\ref{sec:pcl} for detailed derivations.).

\subsection{Photometric Error Estimates}
The radial coordinates from a survey are typically provided as a photometric redshift with some given error, we denote this estimated radial coordinate by 
$\tilde{r}$ and let $r$ represent the true coordinate. Following \cite{Heavens03}, we relate the two coordinates by a conditional probability that we 
model as a Gaussian:

\begin{equation}
 p \left( \tilde{r} | r \right) d \tilde{r} = \frac{1}{\sqrt{2 \pi} \sigma_z} \exp \left[ - 
 \frac{\left( z_{\tilde{r}} - z_r \right)^2}{2 \sigma^2_z} \right] d z_{\tilde{r}}
\end{equation}

\n
where $z_{\tilde{r} , r}$ are the redshifts associated with the given coordinate and $\sigma_z$ is the error. We assume that the error has 
values, $\sigma_z \sim 0.02 - 0.1$ or more and it is important to note that $\sigma_z$ may vary with redshift. 
We can now construct harmonics that represent the average value of the expansion coefficients by 
using the relation between the estimated distance from photometric redshifts, $\tilde{r}$, and the true distance $r$ in terms of the conditional probability:

\begin{equation}
 \Psi_{lm} (k) = \sqrt{\frac{2}{\pi}} \int d^3 \tilde{{\bf{r}}} \int {\bf{r}} \, p ( \tilde{r} | r ) \,  \Psi ({\bf{r}}) \, k \, j_l (k \tilde{r} )  \, 
 Y^{\ast}_{lm} (\hat{\Omega}) . 
\end{equation}

\n
Such a Gaussian error leads to photometric redshift smoothing. 

\subsection{Error Estimate}
The signal to noise for individual modes for a given power-spectrum can be expressed as:
\be
{\delta \myC_{\ell} (k,k) \over \myC_{\ell} (k,k)}\\
 = \sqrt{2 \over 2\ell +1} \left ( 1+ {1 \over \bar n \myC_{\ell} (k,k)} \right )
\ee
Where $\bar n$ is the average number density of galaxies and the second term represents the leading order shot-noise contribution. For our results we take
$\bar n = 10^{-3}h^{3}{\rm Mpc}^{-3}$.

\begin{figure}
\begin{center}
\includegraphics[width=0.35\textwidth]{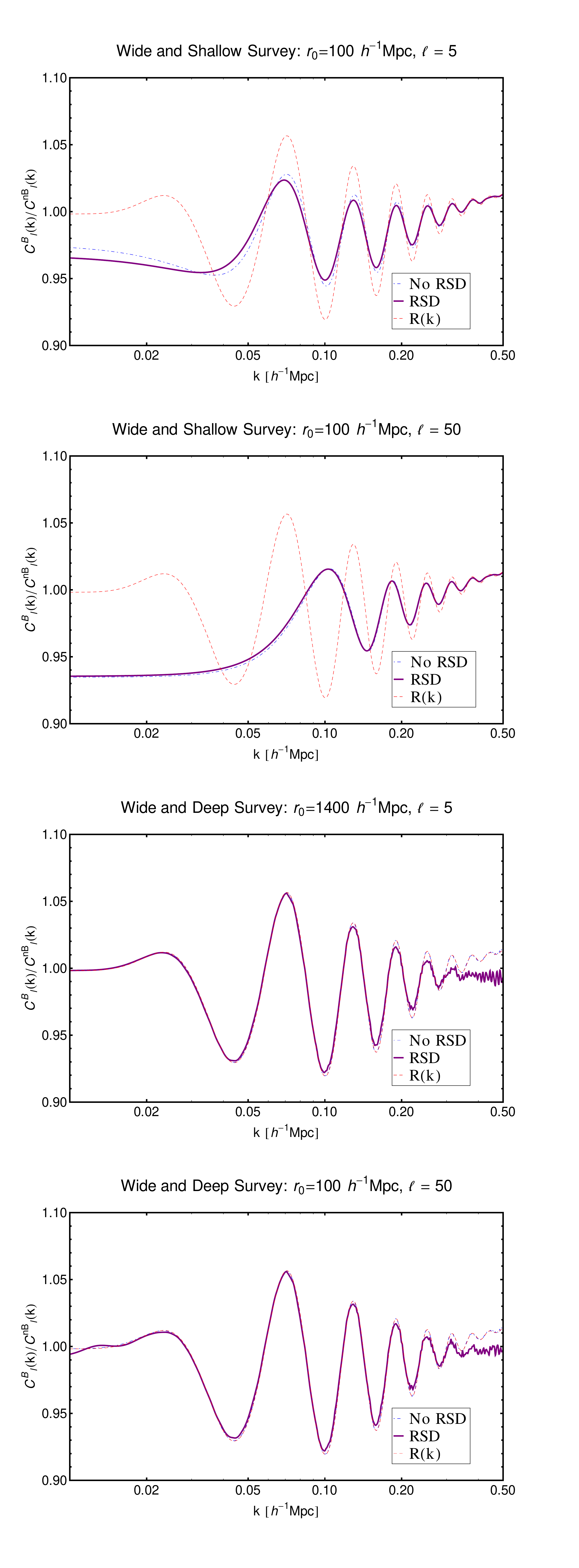}
\end{center}
\caption{Slice in l-space showing $R^{C}_l (k)$ for a wide and shallow survey of $r_0 = 100 h^{-1} Mpc$ at $\ell = 5$ (1st panel) and $\ell = 50$ (2nd panel)
and for a wide and deep survey of $r_0 = 1400 h^{-1} \textrm{Mpc}$ at $\ell = 5$ (3rd panel) and $\ell = 50$ (4th panel). The blue line denotes the 
$\myC^{(00)}$ term, the purple line the sFB spectra incorporating RSD and the red line shows the Fourier space power spectra. In the linear regime the linear prefactors for RSD in the Fourier power spectra cancel and the results correspond to the unredshifted Fourier space power spectra.}
\label{fig:RSD}
\end{figure}

\section{Non-Linear Corrections}
\label{sec:pert}
The role of nonlinear gravitational clustering can investigated in the sFB formalism by incorporating higher-order corrections to the power 
spectrum as described in perturbation theory. The approach we adopt here is standard perturbation theory (SPT), also known as Eulerian perturbation theory, which provides a rigorous framework from which we can 
investigate the the structure of the sFB spectra in a fully analytic manner 
\citep{Vishniac83, Fry84, Goroff86, Suto91, Makino92, Jain94, Scoccimarro96}. Standard perturbation theory is one of the most straightforward approaches 
to studies beyond linear theory and is based on a series solution to the hydrodynamical fluid equations in powers of an initial density or velocity field. 
The nonlinear clustering of matter arises from mode-mode couplings of density fluctuations 
and velocity divergence as seen from the Fourier space equations. The role of perturbation theory in the nonlinear evolution of the BAO in the 
power spectrum has been previously investigated (for an incomplete selection of references please see: 
\cite{Jeong06,Nishimichi07,Nomura08,Nomura09,Taruya09,Taruya10}). 
In this paper we generalise these investigations to the sFB approach. The redshift of the Fourier space power spectra was taken to be $z \sim 0.2$ and the effects of growth have not been analysed in detail. For small surveys the growth does not seem to have significant effects.

\subsection{Standard Perturbation Theory}
Consider the hydrodynamic equations of motion for density perturbations $\delta$ such that our coming coordinates are denoted by ${\bf{x}}$ and the conformal time 
by $\eta$:
\begin{align}
  &\delta^{\prime} ({\bf{x}},\eta) + \nabla \cdot \left[ (1+\delta ({\bf{x}},\eta)) {\bf{v}} ({\bf{x}},\eta) \right] = 0, \\
  &{\bf{v}}^{\prime} ({\bf{x}},\eta) + \left[ {\bf{v}} ({\bf{x}},\eta) \cdot \nabla \right] {\bf{v}} ({\bf{x}},\eta) + \calH (\eta) {\bf{v}} ({\bf{x}},\eta) = - \nabla 
    \phi ({\bf{x}},\eta) , \\
  &\nabla^2 \phi ({\bf{x}},\eta) = \frac{3}{2} \calH^2 (\eta) \delta({\bf{x}},\eta) ,
\end{align}
where a prime denotes the derivative with respect to the conformal time and $\calH = a^{\prime} / a$. The rotational mode of the peculiar velocity ${\bf{v}}$ is a 
decaying solution in an expanding universe and can be neglected in this approach. We introduce a scalar field describing the velocity divergence:
\begin{align}
  \Theta ({\bf{x}}, \eta) = \nabla \cdot {\bf{v}}({\bf{x}},\eta) . 
\end{align}
In our discussion we will focus on a description of the density perturbations and Fourier decompose the above equations to set up and solve a system of 
integro-differential equations. The Fourier decomposition of the perturbations are defined by:
\begin{align}
  \delta ({\bf{x}},\eta) &= \int \frac{d^3 k}{(2 \pi)^3} \delta ({\bf{k}},\eta) e^{-i {\bf{k}} \cdot {\bf{x}}} , \\
  \Theta ({\bf{x}},\eta) &= \int \frac{d^3 k}{(2 \pi)^3} \Theta ({\bf{k}},\eta) e^{-i {\bf{k}} \cdot {\bf{x}}}
\end{align}
The equations of motion can be decomposed as follows:
\begin{align}
&\delta^{\prime} ({\bf{x}},\eta) + \Theta({\bf{k}},\eta) = - \int d^3 k_1 \int d^3 k_2 \delta^{(3)}({\bf{k}}_1 + {\bf{k}}_2 - {\bf{k}}) 
    \nonumber \\ &\qquad \qquad \frac{{\bf{k}} \cdot {\bf{k}}_1}{k^2_1} \Theta ({\bf{k}}_1,\eta) \delta({\bf{k}}_2,\eta), \\
&\Theta^{\prime}({\bf{k}},\eta) + \calH(\eta) \Theta({\bf{k}},\eta) + \frac{3}{2} \calH^2(\eta) \delta({\bf{k}},\eta) = \nonumber \\&  - \int d^3 k_1 \int d^3 k_2 
    \delta^{(3)} ({\bf{k}}_1 + {\bf{k}}_2 - {\bf{k}}) \frac{k^2 ({\bf{k}}_1 \cdot {\bf{k}}_2)}{2 k^2_1 k^2_2} \Theta({\bf{k}}_1,\eta) \Theta({\bf{k}}_2,\eta) . 
\end{align}
In order to solve these coupled integro-differential equations we introduce a perturbative expansion of our variables:
\begin{align}
  & \delta({\bf{k}},\eta) = \displaystyle\sum^{\infty}_{n=1}a^n(\eta) \delta_n ({\bf{k}}) , \\
  & \Theta({\bf{k}},\eta) = \calH(\eta) \displaystyle\sum^{\infty}_{n=1}a^n(\eta) \Theta_n ({\bf{k}}) . 
\end{align}
The general n-th order solutions are given by:
\begin{align}
  &\delta_n ({\bf{k}}) = \int d^3 q_1 \dots \int d^3 q_n \delta^{(3)} \left( \displaystyle\sum^{n}_{i=1} {\bf{q}}_i - {\bf{k}} \right)  
      \nonumber \\ &\qquad \qquad \times F_n({\bf{q}}_1,\dots,{\bf{q}}_n) \displaystyle\Pi^{n}_{i=1} \delta_1 ({\bf{q}}_i) ,  \\
  &\Theta_n ({\bf{k}}) = - \int d^3 q_1 \dots \int d^3 q_n \delta^{(3)} \left( \displaystyle\sum^{n}_{i=1} {\bf{q}}_i - {\bf{k}} \right)  
      \nonumber \\ &\qquad \qquad \times G_n({\bf{q}}_1,\dots,{\bf{q}}_n) \displaystyle\Pi^{n}_{i=1} \delta_1 ({\bf{q}}_i) ,
\end{align}
\noindent
where the kernels $F_n ({\bf{q}}_1 , \dots , {\bf{q}}_n)$ and $G_n ({\bf{q}}_1 , \dots , {\bf{q}}_n)$ are given by \cite{Jain94}:
\begin{align}
  &F_n ({\bf{q}}_1 , \dots , {\bf{q}}_n) = \displaystyle\sum^{n-1}_{m=1} \frac{G_m ({\bf{q}}_1 , \dots , {\bf{q}}_m)}{(2n+3)(n-1)}\nonumber  \\
  &\qquad \nonumber \times \Bigg[ (1+2n) \frac{ {\bf{k}} \cdot {\bf{k}}_1}{k^2_1} F_{n-m} ({\bf{q}}_{m+1} , \dots,{\bf{q}}_n) \nonumber \\
  &\qquad   + \frac{k^2 ({\bf{k}}_1 \cdot {\bf{k}}_2)}{k^2_1 k^2_2} G_{n-m} ({\bf{q}}_{m+1} , \dots,{\bf{q}}_n) \Bigg] ,   \\
  &G_n ({\bf{q}}_1 , \dots , {\bf{q}}_n) = \displaystyle\sum^{n-1}_{m=1} \frac{G_m ({\bf{q}}_1 , \dots , {\bf{q}}_m)}{(2n+3)(n-1)} \nonumber \\
  &\qquad \nonumber \times \Bigg[ 3 \frac{ {\bf{k}} \cdot {\bf{k}}_1}{k^2_1} F_{n-m} ({\bf{q}}_{m+1} , \dots,{\bf{q}}_n) \\
  &\qquad   + n \frac{k^2 ({\bf{k}}_1 \cdot {\bf{k}}_2)}{k^2_1 k^2_2} G_{n-m} ({\bf{q}}_{m+1} , \dots,{\bf{q}}_n) \Bigg] , 
\end{align}
\noindent
The kernel $F_n ({\bf{q}}_1 , \dots , {\bf{q}}_n)$ is not symmetric under permutations of the argument ${\bf{q}}_1 \dots {\bf{q}}_n$ and must be symmetrised:
\begin{align}
  F^{(s)}_n = \frac{1}{n !} \displaystyle\sum_{\textrm{Permutations}} F_n ({\bf{q}}_1 , \dots , {\bf{q}}_n) . 
\end{align}
As an example, the second order symmetrised solution is given by:
\begin{align}
  F^{(s)}_2 (k_1,k_2) = \frac{5}{7} + \frac{2}{7} \frac{({\bf{k}}_1 \cdot {\bf{k}}_2 )^2}{k^2_1 k^2_2} + \frac{({\bf{k}}_1 \cdot {\bf{k}}_2 )}{2} \left( \frac{1}{k^2_1} 
      + \frac{1}{k^2_2} \right) .
\end{align}
The corresponding second order matter power spectrum represents the linear matter power spectrum plus the additional higher-order corrections. This calculation 
is made under the assumption that the first order density perturbations $\delta_1({\bf{k}})$ constitute a Gaussian random field. The power spectrum up to second order 
is given by:
\begin{align}
 P_{\rm{SPT}} (k,z) = D^2(z) P_{\rm{lin}}(k) + D^4(z) P_2(k) ,
\end{align}
where $P_{\rm{lin}}$ is the conventional linear matter power spectrum and the second order correction are given by:
\begin{align}
  P_2 (k) = P_{22} (k) + 2 P_{13}(k) . 
\end{align}
These terms correspond to the contributions to the 4-point correlation function from the (2,2)-order and the (1,3)-order cross-correlations. The explicit form of 
these terms are given by:
\begin{align}
  P_{22} (k) &= 2 \int d^3 q P_{\rm{lin}} (|{\bf{k}} - {\bf{q}}|) \left[ F^s_2({\bf{q}},{\bf{k}}-{\bf{q}})\right]^2 , \\
  P_{13} (k) &= 3 P_{\rm{lin}} (q) \int d^3 q P_{\rm{lin}} (q) F^s_3 ({\bf{q}},-{\bf{q}},{\bf{k}}) 
\end{align}

\n
and the full equations are presented in Appendix C. 

It should be noted that the analytical predictions arising from standard perturbation theory will 
eventually break down as the non-linear terms become dominant over the linear theory predictions. \cite{Jeong06} demonstrated that one-loop standard perturbation 
theory was able to fit N-body simulations to greater than $1 \%$ accuracy when the maximum wave number $k_{1 \%}$ satisfies 
\citep{Taruya09}:

\begin{equation}
 \frac{k^2_{1 \%}}{6 \pi^2} \int^{k_{1 \%}}_0 dq \; P_{\textrm{lin}} (q ; z) = C
\end{equation}
\n
where $C = 0.18$ in standard perturbation theory. SPT theory relies on a straightforward expansion of the set of cosmological hydrodynamical equations and 
the approach has been repeatedly noted as being insufficiently accurate to model and describe the BAOs \citep{Jeong06,Taruya09,Nishimichi09,Carlson09,Taruya10}. 
In particular the amplitude of SPT predicts a monotonical increase with wavenumber that overestimates the amplitude (Figure [\ref{fig:NL}]) with respect to 
N-body simulations \cite{Taruya09}. This is also seen in the full $(k,\ell)$ space spectra in Figures [\ref{fig:C-100-NRSD-NL}-\ref{fig:C-1400-NRSD-NL}]. 

\subsection{Results: SPT}
\begin{figure}
\begin{minipage}[b]{0.5\textwidth}
\begin{center}
\includegraphics[width=1\textwidth]{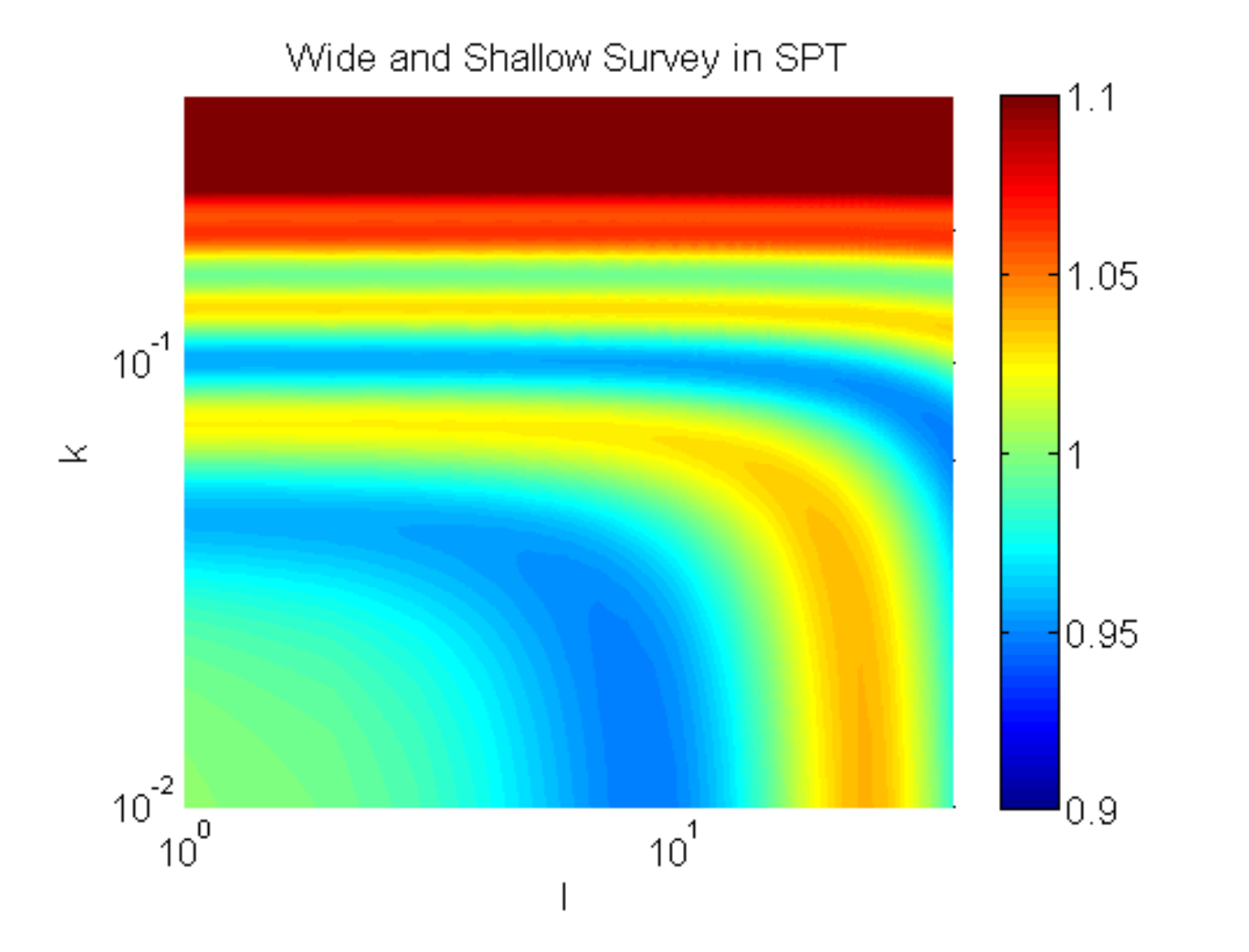}
\end{center}
\caption{Ratio $R^{C}_{\ell} (k)$ of sFB spectrum with and without the physical effects of baryons in $(\ell,k)$ phase space 
for a wide and shallow survey of $r_0 = 100 h^{-1} 
\textrm{Mpc}$ using a Gaussian selection function but with the inclusion of non-linear features as calculated in Standard Perturbation Theory. }
\label{fig:C-100-NRSD-NL}
\end{minipage}

\hspace{0.5cm}

\begin{minipage}[b]{0.5\textwidth}
\includegraphics[width=1\textwidth]{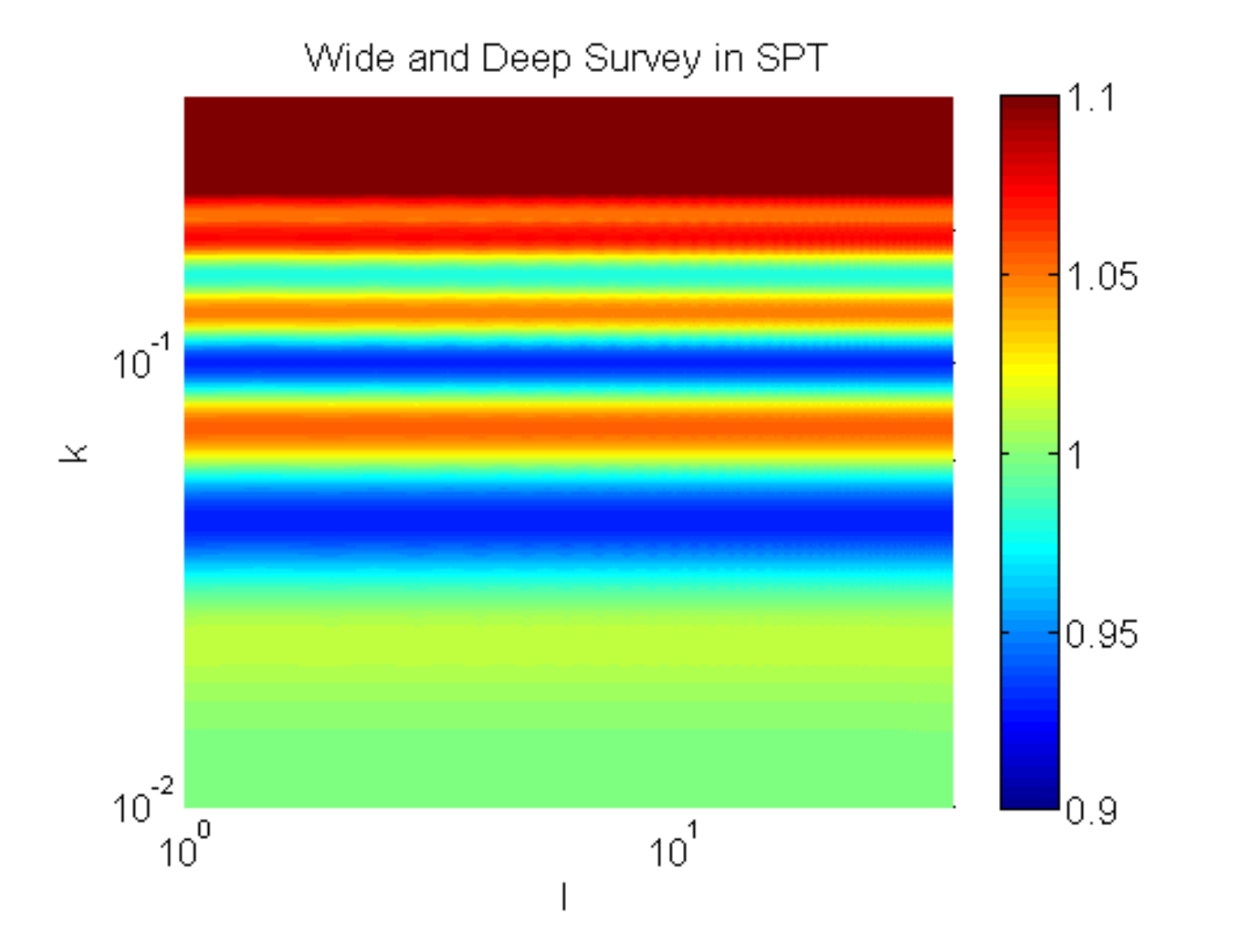}
\begin{center}
\end{center}
\caption{Ratio $R^{C}_{\ell} (k)$ of sFB spectrum with and without the physical effects of baryons in $(\ell,k)$ phase space for a 
wide and deep survey of $r_0 = 1400 h^{-1} 
\textrm{Mpc}$ using a Gaussian selection function but with the inclusion of non-linear features as calculated in Standard Perturbation Theory.}
\label{fig:C-1400-NRSD-NL}{}
\end{minipage} 
\end{figure}

In Figures [\ref{fig:C-100-NRSD-NL}-\ref{fig:C-1400-NRSD-NL}] we have divided the nonlinear power spectrum by a linear no-baryon power spectrum when constructing the 
ratio $R^C_{\ell} (k)$ highlighting the scale dependence introduced by mode coupling. An alternative possibility would 
be to divide the nonlinear 
power spectrum $P^{\textrm{NL}}$ by a power spectrum constructed from smoothing the non-linear spectrum $P^{\textrm{NL}}_{\textrm{smooth}}$ that removes the scale dependence 
and allows for a more detailed comparison of PT predictions 
against numerical simulations. We construct the ratios as follows:

\begin{align}
 R^{C,\textrm{NL/SPT}}_{\ell} (k) &= \frac{C_{\ell}^{\textrm{NL/SPT,B}} (k)}{C_{\ell}^{\textrm{Lin,nB}} (k)} \label{F2Ratio1}\\
 R^{C,\textrm{Lin}}_{\ell} (k) &= \frac{C_{\ell}^{\textrm{Lin,B}} (k)}{C_{\ell}^{\textrm{Lin,nB}} (k)} \label{F2Ratio2}\\
 R^{P,\textrm{NL/SPT}} (k) &= \frac{P^{\textrm{NL/SPT,B}} (k)}{P^{\textrm{Lin,nB}} (k)} \label{F2Ratio3}
\end{align}

\n
In Figure [\ref{fig:NL}] the blue spectra corresponds to Eq.(\ref{F2Ratio1}), the purple spectra to Eq.(\ref{F2Ratio2}) and the red spectra to Eq.(\ref{F2Ratio3}). These spectra do not incorporate RSD. In Figures[\ref{fig:C-100-NRSD-NL}-\ref{fig:C-1400-NRSD-NL}] the ratio Eq.(\ref{F2Ratio1}) is used.

\subsection{Lagrangian Perturbation Theory}
LPT \citep{Matsubara08a} provides a description of the formation of structure by relating the Eulerian coordinates, ${\bf{x}}$, to comoving 
coordinates, ${\bf{q}}$, through the displacement field $\Psi ( {\bf{q}} , t )$:

\begin{equation}
 {\bf{x}} ({\bf{q}} , t) = {\bf{q}} + \Psi ( {\bf{q}} , t) .
\end{equation}

\n
With the assumption that the initial density field is sufficiently uniform, the Eulerian density field $\rho ( {\bf{x}} )$ will satisfy the continuity 
relation $\rho ({\bf{x}}) \, d^3 x = \bar{\rho} \, d^3 q$ where we have denoted the mean density in comoving coordinates by $\bar{\rho}$. The fraction densities 
will then be given by:

\begin{align}
 \delta ({\bf{x}}) &= \int d^3 \, q \, \delta^3 \, \left[ {\bf{x}} - {\bf{q}} - \Psi ( {\bf{q}} ) \right] - 1 , \\
 \delta ({\bf{k}}) &= \int d^3 \, q \, e^{-i {\bf{k}} \cdot {\bf{q}}} \left[ e^{-i {\bf{k}} \cdot \Psi ({\bf{q}}) } - 1 \right] . 
\end{align}

\n
Assuming a pressureless self-gravitating Newtonian fluid in an expanding FLRW universe, the equations of motion for the displacement field are given by 
\cite{Matsubara08a}:

\begin{equation}
 \frac{ d^2}{d t^2} \Psi + 2 H \frac{d}{dt} \Psi = - \nabla_{{\bf{x}}} \phi \left[ {\bf{q}} + \Psi ( {\bf{q}} ) \right] ,
\end{equation}

\n
where $\phi$ is the gravitation potential as determined by Poisson's equation: $\nabla^2_{{\bf{x}}} \phi ({\bf{x}}) = 4 \pi G \bar{\rho} a^2 \delta ( {\bf{x}} )$. LPT 
proceeds by performing a perturbative series expansion of the displacement field:

\begin{align}
 \Psi &= \Psi^{(1)} + \Psi^{(2)} + \cdots \\ 
 \Psi^{(N)} &= \mathcal{O} \left( \left[\Psi^{(1)} \right]^N \right)
\end{align}

\n
The perturbative terms in the series expansion can be written schematically as:

\begin{align}
 \tilde{\Psi}^{(n)} (p) &= \frac{i}{n !} D^n (t) \int \frac{d^3 p_1}{(2 \pi)^3} \cdot \frac{d^3 p_n}{(2 \pi)^3} \delta^3 \left( \displaystyle\sum^n_{j=1} 
 p_j - p \right) \\ \nonumber
 &\qquad \times L^{(n)} ( p_1 , \cdot , p_n ) \delta_0 (p_1) \cdots \delta_0 (p_n) . 
\end{align}

\n
We can perform a similar expansion for both the fractional density and the power spectrum, further details can be found in \cite{Matsubara08a} and 
we will just introduce the results for the power spectrum and how it relates to the predictions of SPT. The power spectrum can be written as:

\begin{equation}
 P(k) = \int d^3 \, q \, e^{-i {\bf{k}} \cdot {\bf{q}} } \left( \left\langle e^{- {\bf{k}} \cdot \left[ \Psi ( {\bf{q}}_1 ) - 
 \Psi ( {\bf{q}}_2 ) \right] } \right\rangle - 1 \right) .
\end{equation}

\n
The two main types of terms that we find in these equations are those terms that depend only on a single position, which are factored out into the first 
exponential term, and those terms that depend on some separation between positions, as seen in the second exponential term. Using the cumulant expansion theorem 
the power spectrum can be written as:

\begin{align}
 P(k) &= \exp \left[ - 2 \displaystyle\sum^{\infty}_{n = 1} \frac{ k_{i_1} \cdot k_{i_{2n}} }{( 2n )!} A^{(2n)}_{i_1 \cdot i_{2n} } \right] \nonumber \\ 
 &\times \int d^3 q e^{-i {\bf{k}} \cdot {\bf{q}} } \left\lbrace exp \left[ \displaystyle\sum^{\infty}_{N = 2} \frac{ k_{i_1} \cdot k_{i_N} }{(N!)} B^{(N)}_{i_1 
 \cdot i_N} (q) \right] - 1 \right\rbrace 
\end{align}

\n
where $A^{(2n)}_{i_1 \cdot i_{2n}}$ and $B^{(N)}_{i_1 \cdot i_N}$ are given in \cite{Matsubara08a}. $A^{(N)}$ relates to the cumulant of a displacement 
vector at a single position and $B^{(N)}$ relates to the cumulant of two displacement vectors separated by $| {\bf{q}} |$. Expanding both the $A^{(N)}$ and 
the $B^{(N)}$ terms yields SPT. \cite{Matsubara08a}, however, proposes expanding only the $B^{(N)}$ terms and leaving the $A^{(N)}$ terms as an exponential 
prefactor. The justification for this is that this exponential prefactor will contain infinitely higher-order perturbations in terms of SPT and has effectively 
given a way to resum the infinite series of perturbations found in SPT. Expanding and solving for the $B^{(N)}$ terms yields the standard LPT results 
\cite{Matsubara08a}:

\begin{align}
 P(k) &= e^{- (k \Sigma)^2 / 2} \left[ P_{\textrm{lin}} (k) + P_{22} (k) + P^{\textrm{LPT}}_{13} (k) \right] .
\end{align}

\n
The term $P_{22}$ is identical to it's SPT counterpart but the term $P^{\textrm{LPT}}_{13}$ is now slightly modified but retains much of the structure found in 
SPT. 

\subsection{Results: LPT}
\begin{figure}
\begin{minipage}[b]{0.5\textwidth}
\begin{center}
\includegraphics[width=1\textwidth]{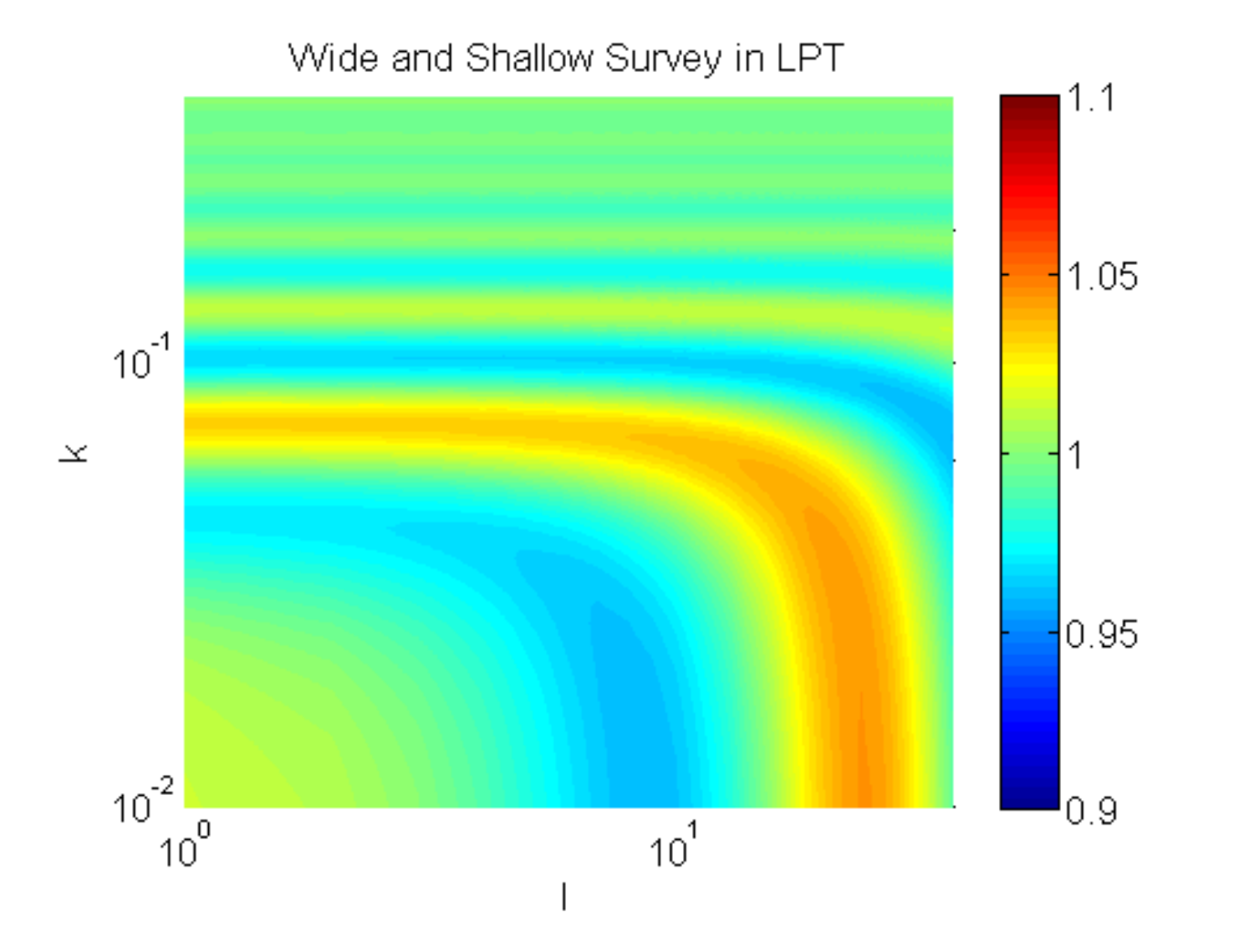}
\end{center}
\caption{Ratio $R^{C}_{\ell} (k)$ of sFB spectrum with and without the physical effects of baryons in $(\ell,k)$ phase space 
for a wide and shallow survey of $r_0 = 100 h^{-1} 
\textrm{Mpc}$ using a Gaussian selection function but with the inclusion of non-linear features as calculated in LPT. }
\label{fig:C-100-NRSD-LNL}
\end{minipage}

\hspace{0.5cm}

\begin{minipage}[b]{0.5\textwidth}
\includegraphics[width=1\textwidth]{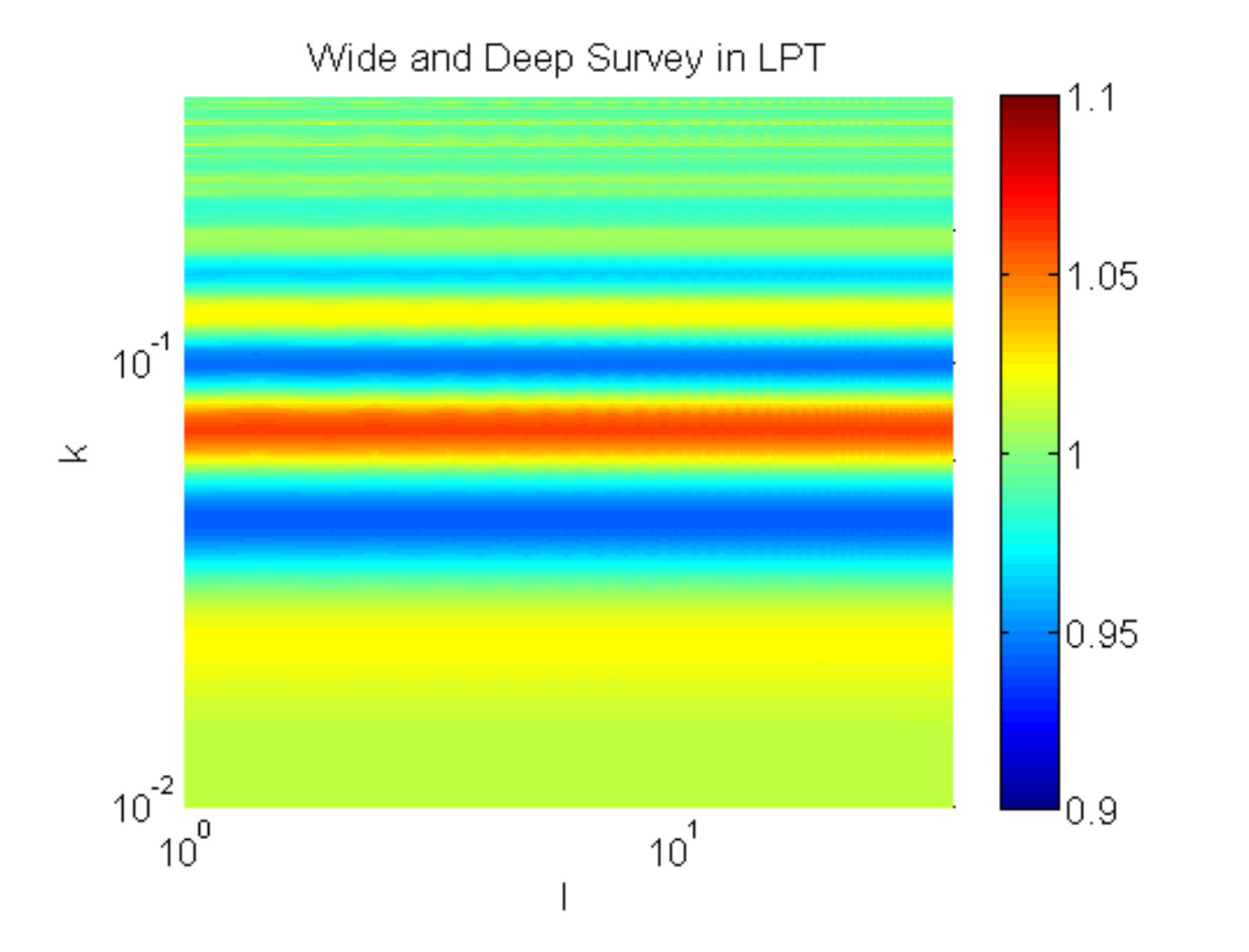}
\begin{center}
\end{center}
\caption{Ratio $R^{C}_{\ell} (k)$ of sFB spectrum with and without the physical effects of baryons in $(\ell,k)$ phase space for a 
wide and deep survey of $r_0 = 1400 h^{-1} 
\textrm{Mpc}$ using a Gaussian selection function but with the inclusion of non-linear features as calculated in LPT.}
\label{fig:C-1400-NRSD-LNL}{}
\end{minipage} 
\end{figure}

In Figures [\ref{fig:C-100-NRSD-LNL}-\ref{fig:C-1400-NRSD-LNL}] we again divide the nonlinear power spectrum by a linear no-baryon power spectrum when constructing the 
ratio $R^C_{\ell} (k)$. The explicit ratios used are:

\begin{align}
 R^{C,\textrm{NL/LPT}}_{\ell} (k) &= \frac{C_{\ell}^{\textrm{NL/LPT,B}} (k)}{C_{\ell}^{\textrm{Lin,nB}} (k)} \label{F3Ratio1}\\
 R^{C,\textrm{Lin}}_{\ell} (k) &= \frac{C_{\ell}^{\textrm{Lin,B}} (k)}{C_{\ell}^{\textrm{Lin,nB}} (k)} \label{F3Ratio2}\\
 R^{P,\textrm{NL/LPT}} (k) &= \frac{P^{\textrm{NL/LPT,B}} (k)}{P^{\textrm{Lin,nB}} (k)} \label{F3Ratio3}
\end{align}

\begin{figure}
\begin{center}
\includegraphics[width=0.355\textwidth]{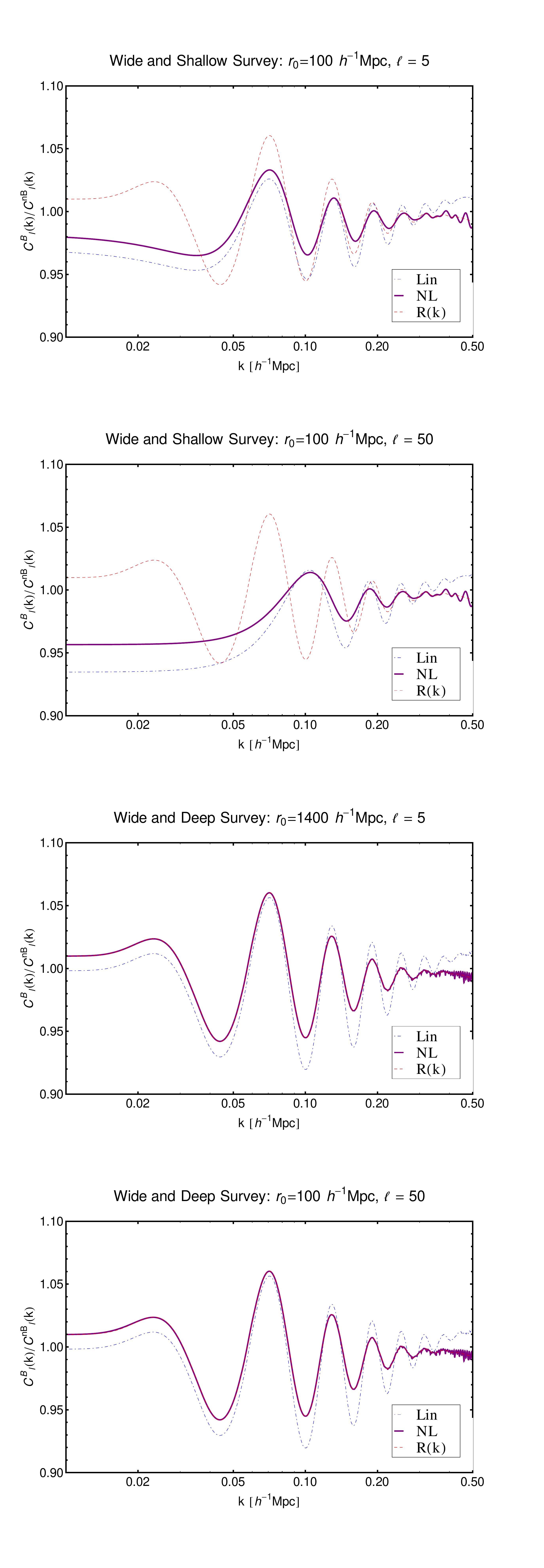}
\end{center}
\caption{Slice in l-space showing $R^{C}_{\ell} (k)$ for $\ell = 5$ (top-panels) 
and $\ell = 50$ (bottom-panels) 
in a wide and shallow survey of $r_0 = 100 h^{-1} \textrm{Mpc}$ (left-panels) and 
a wide and deep survey of $r_0 = 1400 h^{-1} \textrm{Mpc}$ (right-panels). The solid blue line denotes the linear results, the solid 
purple line the non-linear 1-loop LPT spectra and the dashed line the non-linear 1-loop LPT spectra. LPT consistently underestimates 
the power spectrum in the large-k limit contrasting to the divergence at large-k in 1-loop SPT results. This 
difference occurs due to the effective ressumation of an infinite series of perturbations from SPT that occurs in LPT. }
\label{fig:LNL}
\end{figure}

\n
In Figure [\ref{fig:LNL}] the blue spectra corresponds to Eq.(\ref{F3Ratio1}), the purple spectra to Eq.(\ref{F3Ratio2}) and the red spectra to Eq.(\ref{F3Ratio3}). These spectra do not incorporate RSD. In Figures[\ref{fig:C-100-NRSD-LNL}-\ref{fig:C-1400-NRSD-LNL}] the ratio Eq.(\ref{F3Ratio1}) is used.

The sFB can be seen to mimic the predictions of LPT in consistently underestimating the power at large k but we also see that the sFB power spectra radialise 
towards the non-linear LPT spectra in the limit $r \rightarrow \infty$. This can be seen in Figure [\ref{fig:LNL}] where the non-linear sFB tends towards the Fourier space power spectrum in amplitude and phase. We have included a comparison to the linear sFB spectra, which we know to radialise to the linear Fourier space spectra.
This behaviour is completely expected due to the nature of the sFB formalism and the fact that 
the resulting angular spectra are still constructed via products of Bessel functions which form an orthogonal set of basis functions. As such we do not observe 
the types of mode-mixing that are inherent when considering RSD in the sFB formalism. The damping and smearing of the BAOs in this instance is purely 
from gravitational instability and is encapsulated in the power spectrum. We also note that the full $(\ell , k)$ plane is an interesting arena for 
visualising some of the differences in behaviour between various models for structure formation. This can be seen in the changes to the widths and amplitudes 
of the BAO wiggles as seen in the plane in Figures [\ref{fig:C-100-NRSD-NL}-\ref{fig:C-1400-NRSD-LNL}]. 

As future wide field surveys will cover both wide and deep regions of the sky we can use the sFB formalism as a tool to distinguish between different models for 
non-linear evolution of the matter density field. Interesting questions include, how do different theories affect the distribution of 
power in the radial and tangential modes? How can the sFB formalism be expanded to compare the RSD results to those as derived from 
higher-order perturbation theory? How can we best characterise 
the sFB spectra and how can we characterise the radialisation of information in these higher-order models? The analysis and results to these questions 
will be presented in a forthcoming paper. 

\begin{figure}
\begin{center}
\includegraphics[width=0.355\textwidth]{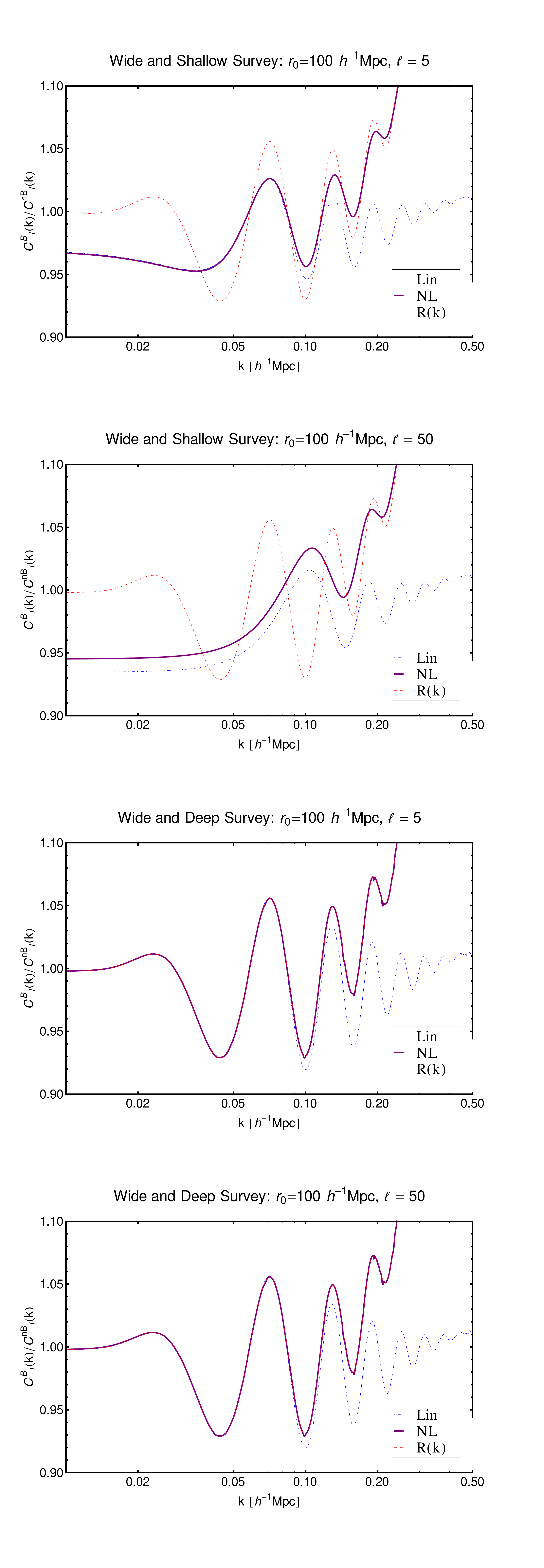}
\end{center}
\caption{Slice in l-space showing $R^{C}_{\ell} (k)$ for $\ell = 5$ (top-panels) 
and $\ell = 50$ (bottom-panels) 
in a wide and shallow survey of $r_0 = 100 h^{-1} \textrm{Mpc}$ (left-panels) as well as for
a deep survey of $r_0 = 1400 h^{-1} \textrm{Mpc}$ (right-panels). The solid blue line represents the linear angular spectra, the solid 
purple line the non-linear 1-loop SPT angular spectra and the dashed line the non-linear 1-loop SPT power spectrum. SPT consistently overestimates 
the linear power spectrum in the large-k limit and it is well known that SPT works well at high-$z$ and large scales.}
\label{fig:NL}
\end{figure}

\section{Results}
\label{sec:res}
Following \cite{Rassat12} we construct the quantity $R^{C}_{\ell} (k)$ to isolate the BAOs in the sFB formalism. 
The matter power spectrum includes the physical effects of baryons leading to the characteristic oscillations as seen in 
Fourier space \citep{Sunyaev70,Peebles70,Seo03,Seo07}. In our analysis, we have 
adopted the zero-Baryon transfer function of \cite{Eisenstein05} to model the power spectra excluding the physical effects of baryons. 

In Figure [\ref{fig:RSD}] 
we construct slices of constant $\ell$ through $R^{C}_{\ell} (k)$ to investigate how RSD manifest themselves in the oscillations. \cite{Rassat12} used 
such slice plots to investigate the radialisation of information when varying levels of tangential and radial information is included in a survey. 
The radialisation of information can be investigated by notion that in the limit $r_0 \rightarrow \infty$ we find: 

\begin{equation}
\label{eqn:ratioP}
 \lim_{r_0 \rightarrow \infty} R^C_{\ell} (k) = R^P (k) = \frac{P^B (k)}{P^{nB} (k)} . 
\end{equation}

\n
Using this definition, radialisation means that $R^C_{\ell} (k)$ tends towards $R^P (k)$ in both phase and amplitude. 
This occurs as the tangential modes are attenuated due 
to mode-canceling along the line of sight \citep{Rassat12}. 
The radialisation can be seen in Figures [\ref{fig:RSD}-\ref{fig:LNL}]  as the amplitude and phase of the sFB spectra tends towards those of the Fourier space spectra. Additionally the BAOs appear to only have a radial ($k$) dependence in surveys with a large radial parameter $r_0$, as can be seen by the invariance the BAOs under a varying multipole $\ell$. The addition of RSD does not change this trend drastically though we do see more prominent radial and tangential dependence in Figure [\ref{fig:RSD}] with the rate at which the BAOs radialise being affected due to mode-mixing that leads to attenuation and peak shifts. The results appear to be in agreement with previous studies with percent level shifts in the peaks to smaller $k$ and damping of the amplitude \citep{Nishimichi07,Nomura08,Smith08,Nomura09,Taruya10}. As can be seen in Figure [\ref{fig:RSD}] the BAOs seem to effectively radialise, even in the presence of RSD, at large values of the radius parameter $r_0$ and for higher multipoles $\ell$. 
Effective radialisation simply means that the behaviour (i.e. amplitude and phase of the peaks and troughs) of the sFB spectra with RSD asymptotes towards the Fourier space spectra with RSD under the caveat that the intrinsic mode-mixing causes some smearing of radial information and leads to the distinctive damping features seen at high-$k$. The radialisation of information can linked with the preservation of the orthogonality of the basis functions. In the case of RSD, the appearance of derivatives of spherical Bessel functions guarantees that the basis will not be perfectly orthogonal and we observe mode-mode coupling and the generation of off-diagonal contributions. In higher-order PT, the basis functions are still spherical Bessel functions and we observe the radialisation as per linear theory. The behaviour of the non-linear sFB spectra in the full $(\ell , k)$ space is naturally different for various descriptions of non-linearity in gravitational collapse. 

\begin{figure}
\begin{minipage}[b]{0.5\textwidth}
\begin{center}
\includegraphics[width=1\textwidth]{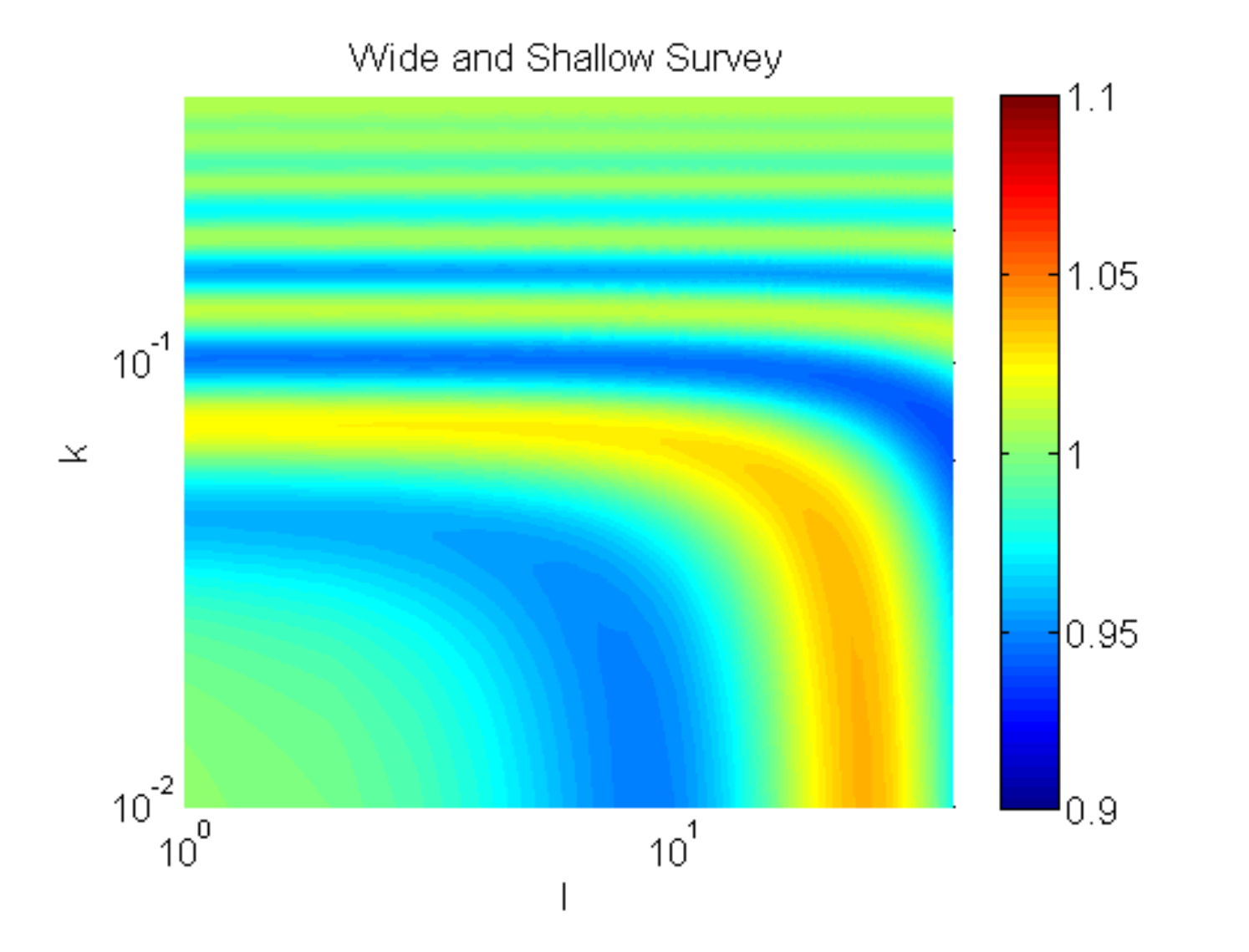}
\end{center}
\caption{Ratio $R^{C}_{\ell} (k)$ of sFB spectrum with and without the physical effects of baryons in $(\ell,k)$ phase 
space for a wide and shallow survey of $r_0 = 100 h^{-1} 
\textrm{Mpc}$ using a Gaussian selection function. The baryonic wiggles are seen in both the radial (k) and tangential ($\ell$) directions.}
\label{fig:C-100-NRSD-L}
\end{minipage}

\hspace{0.5cm}

\begin{minipage}[b]{0.5\textwidth}
\includegraphics[width=1\textwidth]{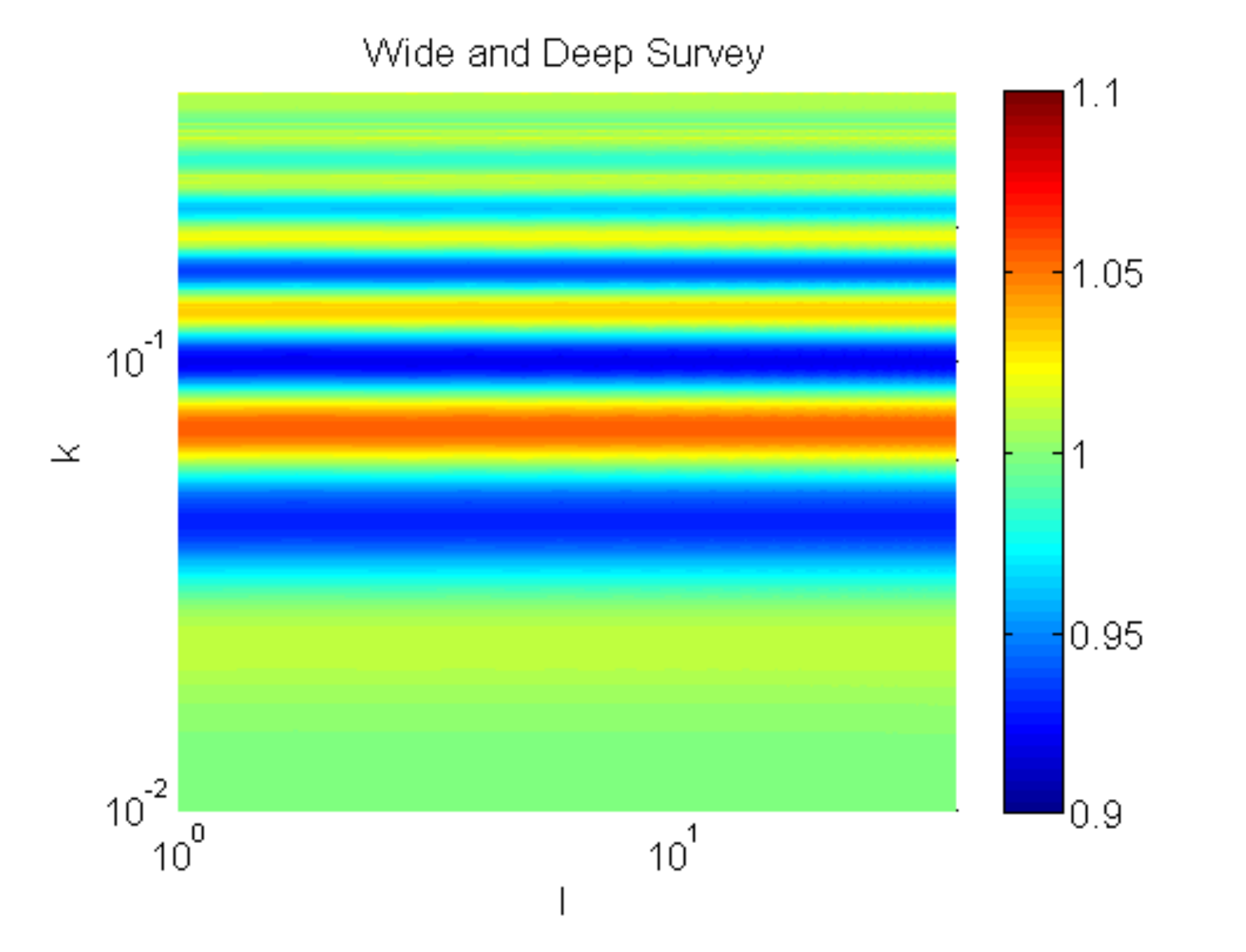}
\begin{center}
\end{center}
\caption{Ratio $R^{C}_{\ell} (k)$ of sFB spectrum with and without the physical effects of baryons in $(\ell,k)$
phase space for a wide and deep survey of $r_0 = 1400 h^{-1} 
\textrm{Mpc}$ using a Gaussian selection function. The baryonic wiggles are seen in both the radial (k) and tangential ($\ell$) directions.}
\label{fig:C-1400-NRSD-L}{}
\end{minipage} 
\end{figure}

\begin{figure}
\begin{minipage}[b]{0.5\textwidth}
\begin{center}
\includegraphics[width=1\textwidth]{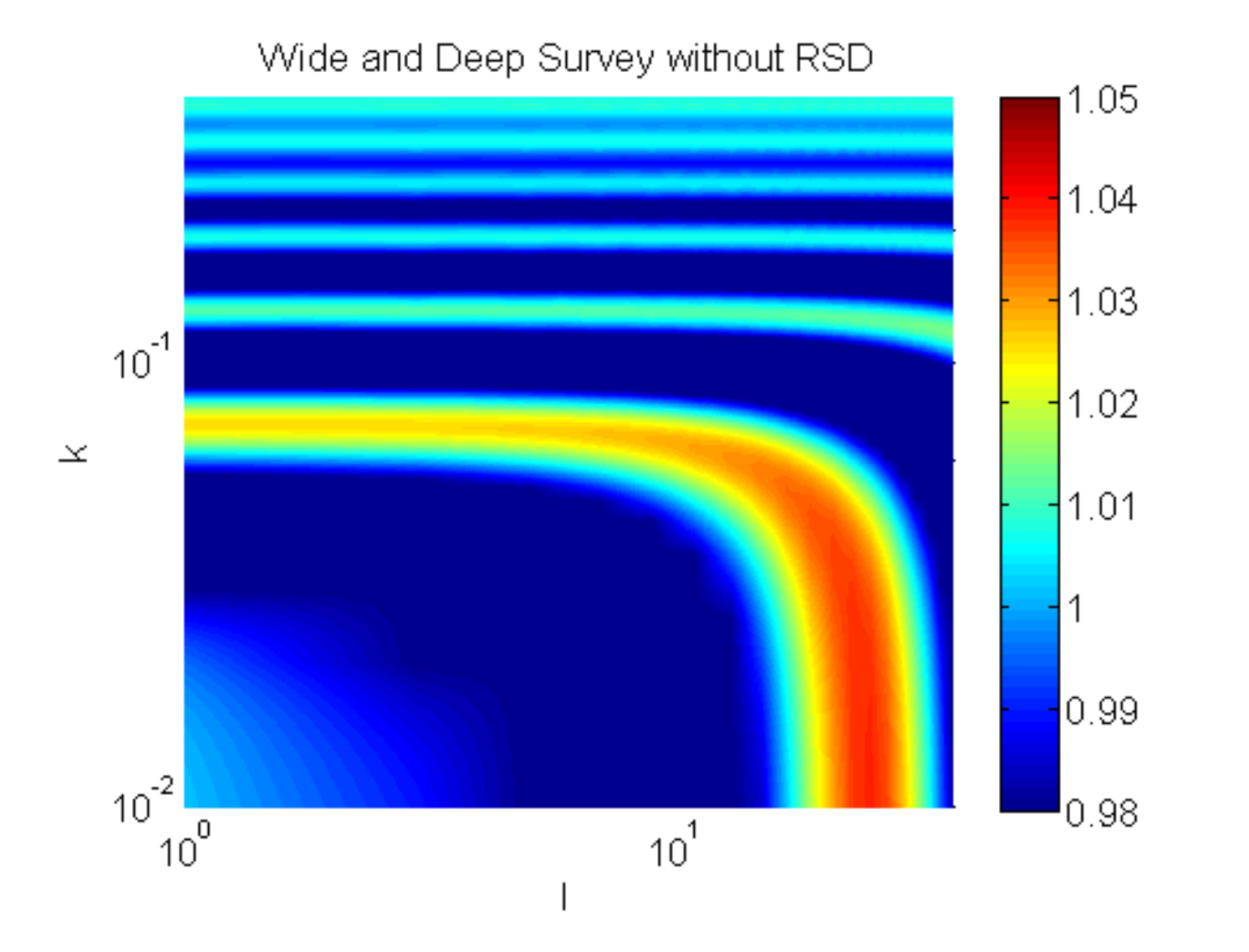}
\end{center}
\caption{Ratio $R^{C}_{\ell} (k)$ of sFB spectrum for a 
wide and shallow survey of $r_0 = 100 h^{-1} 
\textrm{Mpc}$ using a Gaussian selection function without RSD. Here we have reduced the dynamic range to highlight the impact that RSD have on the BAOs. This plot is equivalent to Figure [\ref{fig:C-100-NRSD-L}]. Compare to Figure [\ref{fig:C-100-RSD-L-LDR}] to see the phenomenological effects of RSD. } 
\label{fig:C-100-NRSD-L-LDR}
\end{minipage}

\hspace{0.5cm}

\begin{minipage}[b]{0.5\textwidth}
\includegraphics[width=1\textwidth]{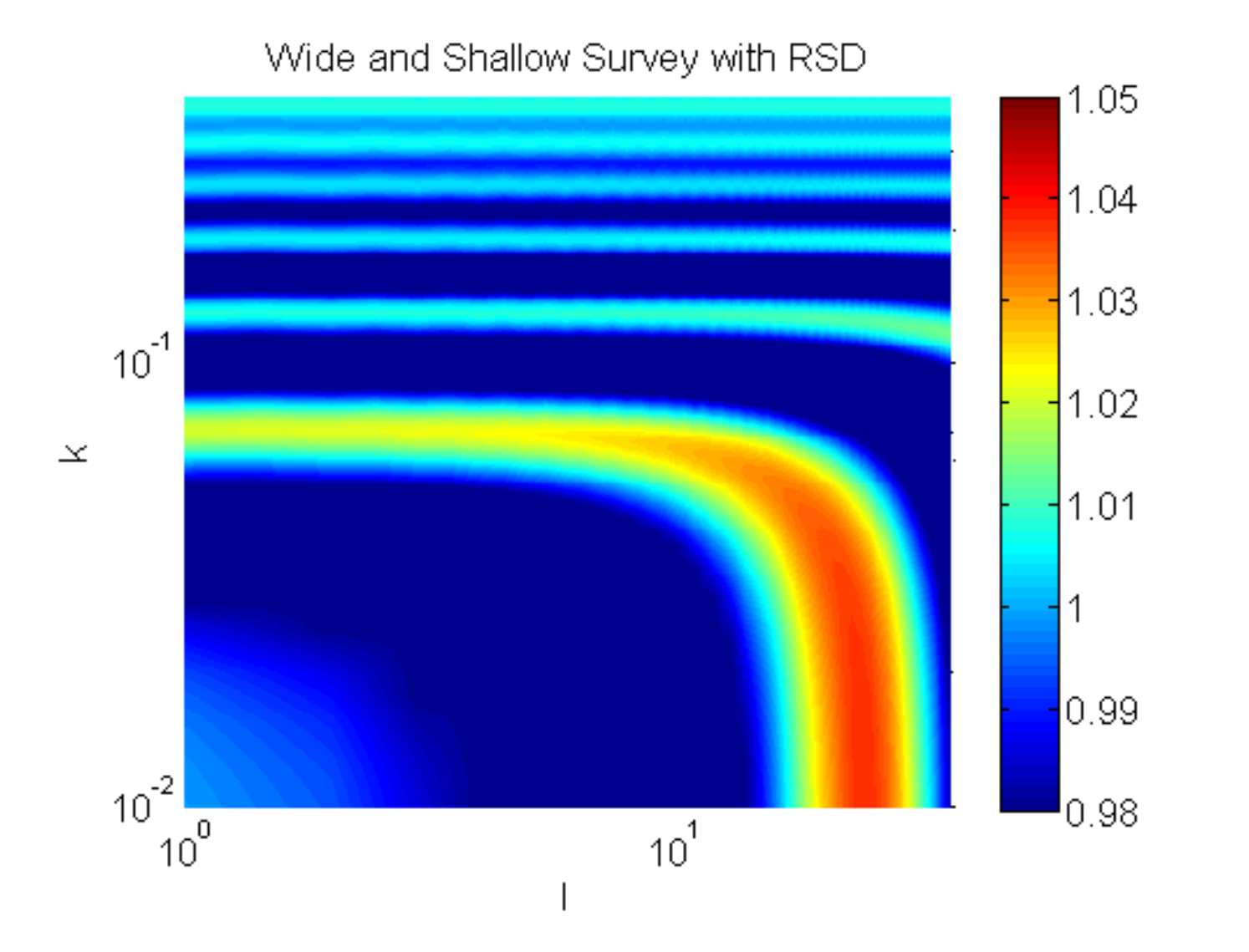}
\begin{center}
\end{center}
\caption{Ratio $R^{C}_{\ell} (k)$ of sFB spectrum for a 
wide and shallow survey of $r_0 = 100 h^{-1} 
\textrm{Mpc}$ using a Gaussian selection function but with the inclusion of RSD. Here we have reduced the dynamic range to highlight the impact that RSD have on the BAOs. This plot is equivalent to Figure [\ref{fig:C-100-RSD}]. Compare to the unredshifted results of Figure [\ref{fig:C-100-NRSD-L-LDR}]. RSD suppress the power at lower $\ell$ and $k$ modes and smear the wiggles in the $k$ direction. Power in the first peak is reduced as per Figure 1 but the amplitudes level at higher $\ell$.}
\label{fig:C-100-RSD-L-LDR}{}
\end{minipage} 
\end{figure}

The BAOs in the sFB formalism will radialise as the survey size, $r_0$, is allowed to increase. This corresponds to the amplitude and the phase of the BAOs tending 
towards the values as measured in the Fourier space ratio $R^P (k)$. As noted in \cite{Rassat12}, for a wide-field shallow survey the BAO will 
have smaller amplitudes and are spread across the $( \ell ,k)$ space. It was also shown in \cite{Rassat12} that the BAOs appear 
to radialise before the full sFB spectrum is able to and notably so at large $\ell$ (Figures [\ref{fig:C-100-NRSD-L}-\ref{fig:C-1400-NRSD-L}]) and this is one of the key motivations for implementing the sFB formalism. With the addition of RSD, the radialisation of information is intrinsically limited due to mode-mixing but a lot of the same phenomenological behaviour can be seen: dependence on radial modes and not on tangential modes at large $r$ and the asymptotic behaviour toward the Fourier space spectra at large $r$. 

\n

\section{Conclusion}
\label{sec:conclu}
The baryon acoustic oscillations give rise to a characteristic signature in the observed matter power spectrum that acts as a standard ruler. 
Unfortunately, the observed matter 
power spectrum is contaminated and complicated by the non-linear evolution of density perturbations, galaxy clustering bias, RSD and survey specific systematic errors. Additionally, upcoming future surveys will cover both large and deep areas of the sky 
demanding a formalism that simultaneously treats the both the spherical sky geometry and the extended radial coverage. The sFB basis 
was proposed as a natural basis for random fields in this geometry. The recent study by \cite{Rassat12} 
was an initial step into investigating the role of the sFB formalism in the study and analysis of the BAO. This study, however, did not go as far as including 
higher-order contributions to the power spectrum that may impact the radialisation of information by introducing, for example, mode-mode couplings. The 
stability of this radialisation of information and the information content of tangential ($\ell$) and radial ($k$) modes for higher-order physics is the 
key topic of interest. 

In this paper we have presented a short treatment of the effects of {\bf linear} RSD and non-linear corrections to measurements of 
baryon acoustic oscillations in the sFB expansion. In order to guide this investigation we have extended the formalism and techniques 
outlined in \cite{Rassat12} and the appropriate machinery for partial-sky coverage was introduced. In particular we have been able to use the procedure outlined in \cite{Heavens95} to construct a series expansion solution to  
model RSD. This solution was used to numerically and analytically investigate the modulation to the angular 
sFB power spectrum. The qualitative behaviour of these 
corrections was outlined for surveys with varying levels of radial ($k$-modes) and tangential ($\ell$-modes) information. It was seen that the RSD impact the 
radialisation of information through mode-mixing that generates a distinct signature in the spectra. These RSD were investigated over a range of survey configurations. The mode-mode coupling was related to the presence of derivatives of spherical Bessel functions and was contrasted to the linear Kaiser result in which the basis functions are constructed from plane waves or derivatives of plane waves which simply return a plane wave of the same frequency and preserve orthogonality. This mode-mode coupling can therefore be thought of as a geometrical artifact in the sFB formalism arising from RSD on large scales \cite{Heavens95,Zaroubi96,Shapiro11}. 

Additionally we considered the structure and form of the sFB spectra when non-linearity arising from gravitational clustering was considered. 
We primarily investigated one-loop corrections to the matter power spectrum arising from two mainstream models for leading order corrections 
as given by SPT and LPT. A brief outline 
of perturbation theory methods was given and the basic equations for SPT and LPT introduced. The non-linear corrections, and 
how we expect them to be independent of the notion of radialisation of information in the BAOs, was numerically investigated. The redshift of the Fourier space power spectrum was taken to be $z \sim 0.2$ and the detailed study of the non-linear corrections with redshift will be presented elsewhere. These are not thought to be important at low redshifts or shallow surveys where the impact of growth seems negligible.

In this paper we have neglected other contributions to the power spectrum such as General Relativistic corrections, lensing terms, 
the role of non-linearities through more detailed studies, more complex treatments of galaxy biasing and more detailed modelling of the 
hydrodynamical and radiative processes involved in these processes \citep{Guillet10,Juszkiewicz12}. In addition we have not considered the role of systematic errors 
associated with a given survey. It would be interesting to compare the results from SKA-like configurations and N-body simulations but we leave this to a future paper. 

\section{Acknowledgements}
\label{acknow} 
DM acknowledges support
from STFC standard grant ST/G002231/1 at the School of Physics and
Astronomy at Cardiff University where this work was completed. 
Its a pleasure to thanks Peter Coles and Alan Heavens, 
for many useful discussions. 
We note that a paper by \cite{Yoo13} appeared 
on the arXiv shortly after submission of this paper.

\bibliography{Bao.bbl}

\begin{thebibliography}{99}
\bibitem[\protect\citeauthoryear{Abramo, Reimberg and Xavier}{2010}]{Abramo10} Abramo L. R., Reimberg P.  H., Xavier H. S., 2010, PRD, 82, 043510, arXiv:1105.0563 
\bibitem[\protect\citeauthoryear{Adelman-McCarthy et al}{2008}]{Adelman08} Adelman-McCarthy J. K., Ag\"ueros M. A., Allam S. S., et al., 2008, ApJS, 175, 297
\bibitem[\protect\citeauthoryear{Alam and Sahni}{2006}]{Alam06} Alam U., Sahni V., 2006, PRD, 73, 084024, arXiv:astro-ph/0511473
\bibitem[\protect\citeauthoryear{Alcock and Paczynski}{1979}]{Alcock79} Alcock C., Paczynski B., 1979, Nat., 281, 358
\bibitem[\protect\citeauthoryear{Amendola, Quercellini and Giallongo}{2005}]{Amendola05} Amendola L., Quercellini C., Giallongo E., 2005, MNRAS, 357, 429,  	arXiv:astro-ph/0404599
\bibitem[\protect\citeauthoryear{Annis et al}{2005}]{Annis05} Annis J., et al., The Dark Energy Task Force, 2005, arXiv:astro-ph/0510195 
\bibitem[\protect\citeauthoryear{Asorey et al}{2012}]{Asorey12} Asorey J., Crocce M., Gaztanaga E., Lewis A., 2012, arXiv:1207.6487
\bibitem[\protect\citeauthoryear{Banerji, Abdalla, Lahav and Lin}{2008}]{Banerji08} Banerji M., Abdalla F. B., Lahav O., Lin H., 2008, MNRAS, 386, 1219,  	arXiv:0711.1059v2
\bibitem[\protect\citeauthoryear{Benítez et al}{2009}]{Benitez09} Ben\'itez N., et al., 2009, APJ, 691, 241, arXiv:0807.0535v4
\bibitem[\protect\citeauthoryear{Bianchi et al}{2012}]{Bianchi12} Bianchi D., Guzzo L., Branchini E., Majerotto E., de la Torre S., Marulli F., Moscardini L., Angulo R. E., 2012, MNRAS, 427, 2420-2436
\bibitem[\protect\citeauthoryear{Binney and Quinn}{1991}]{Binney91} Binney J., Quinn T., 1991, MNRAS, 241, 678 
\bibitem[\protect\citeauthoryear{Blake et al}{2006}]{Blake06} Blake C., Parkinson D., Bassett B., Glazebrook K., Kunz M., Nichol R. C., 2006, MNRAS, 365, 255,  	arXiv:astro-ph/0510239
\bibitem[\protect\citeauthoryear{Carlson, White and Padmanabhan}{2009}]{Carlson09} Carlson J., White M., Padmanabhan N., 2009, PRD, 80, 043531,  	arXiv:0905.0479
\bibitem[\protect\citeauthoryear{Castro, Heavens and Kitching}{2005}]{Castro05} Castro P. G., Heavens A. F., Kitching T. D., 2005, PRD, 72, 023516,  	arXiv:astro-ph/0503479
\bibitem[\protect\citeauthoryear{Challinor and Lewis}{2011}]{Challinor11} Challinor A., Lewis A., 2011, PRD, 84, 043516, arXiv:1105.5292
\bibitem[\protect\citeauthoryear{Coles and Erdogdu}{2007}]{Coles07} Coles P., Erdogdu P., 2007, JCAP, 0710, 007, arXiv:0706.0412
\bibitem[\protect\citeauthoryear{Colless et al}{2003}]{Colless03} Colless M., et al., 2003, arXiv:astro-ph/0306581, "The 2dF Galaxy Redshift Survey: Final Data Release"
\bibitem[\protect\citeauthoryear{Crocce and Scoccimarro}{2008}]{Crocce08} Crocce M., Scoccimarro R., 2008, PRD, 77, 023533
\bibitem[\protect\citeauthoryear{Crocce, Fosalba, Castander and Gaztanaga}{2010}]{Crocce10} Crocce M., Fosalba P., Castander F. J., Gaztanaga E., 2010, MNRAS, 403, 1353, arXiv:0907.0019
\bibitem[\protect\citeauthoryear{Dolney, Jain and Takada}{2006}]{Dolney06} Dolney D., Jain B., Takada M., 2006, MNRAS, 366, 884-898, arXiv:astro-ph/0409445 
\bibitem[\protect\citeauthoryear{Eisenstein et al}{2005}]{Eisenstein05} Eisenstein D. J., Zehavi I., Hogg D. W., Scoccimarro R., et al., 2005, APJ, 633, 560, arXiv:astro-ph/0501171
\bibitem[\protect\citeauthoryear{Erdogdu et al}{2006}]{Erdogdu06} Erdogdu P., et al., 2006, MNRAS, 373, 45-64, arXiv:astro-ph/0610005
\bibitem[\protect\citeauthoryear{February, Clarkson and Maartens}{2012}]{February12} February S., Clarkson C., Maartens R., 2012, arXiv:1206.1602 
\bibitem[\protect\citeauthoryear{Fisher et al}{1994}]{Fisher94} Fisher K. B., Scharf C. A., Lahav O., 1994, MNRAS, 266, 219-226
\bibitem[\protect\citeauthoryear{Fisher et al}{1995}]{Fisher95} Fisher K. B., Lahav O., Hoffman Y., Lynden-Bell D., Zaroubi S., 1995, MNRAS, 272, 885, arXiv:astro-ph/9406009 
\bibitem[\protect\citeauthoryear{Fry}{1984}]{Fry84} Fry J. N., 1984, ApJ, 279, 499
\bibitem[\protect\citeauthoryear{Heavens}{2003}]{Heavens03} Heavens A., 2003, MNRAS, 343, 1327 
\bibitem[\protect\citeauthoryear{Garcia-Bellido and Haugboelle}{2008}]{Garcia08} Garcia-Bellido J., Haugboelle T., 2008, JCAP, 0804, 003, arXiv:0802.1523 
\bibitem[\protect\citeauthoryear{Garcia-Bellido and Haugboelle}{2009}]{Garcia09} Garcia-Bellido J., Haugboelle T., 2009, JCAP, 0909, 028, arXiv:0810.4939 
\bibitem[\protect\citeauthoryear{Gaztanaga et al}{2011}]{Gaztanaga11} Gaztanaga E., et al., 2011, arXiv:1109.4852
\bibitem[\protect\citeauthoryear{Goroff, Grinstein, Rey and Wise}{1986}]{Goroff86} Goroff M., Grinstein B., Rey S. -J., Wise M., 1986, ApJ, 311, 6
\bibitem[\protect\citeauthoryear{Goobar, Hannestad, M\"ortsell and Tu}{2006}]{Goobar06} Goobar A., Hannestad S., M\"ortsell E., Tu H., 2006, JCAP, 6, 19, arXiv:astro-ph/0602155
\bibitem[\protect\citeauthoryear{Guillet et al}{2010}]{Guillet10} Guillet T., Teyssier R., Colombi S., 2010, MNRAS, 405, 525, arXiv:0905.2615
\bibitem[\protect\citeauthoryear{Guzzo et al}{2008}]{Guzzo08} Guzzo L., et al., 2008, Nature, 451, 541-545
\bibitem[\protect\citeauthoryear{Heavens and Taylor}{1995}]{Heavens95} Heavens A., Taylor A., 1995, MNRAS, 275, 483, arXiv:astro-ph/9409027 
\bibitem[\protect\citeauthoryear{Heavens}{2003}]{Heavens03} Heavens A., 2003, MNRAS, 343, 1327-1334, arXiv:astro-ph/0304151
\bibitem[\protect\citeauthoryear{Hirata}{2009}]{Hirata09} Hirata C., 2009, MNRAS, 399, 1074, arXiv:0903.4929 
\bibitem[\protect\citeauthoryear{Hivon, Bouchet, Colombi, Juszkiewicz}{1995}]{Hivon95} Hivon E., Bouchet F. R., Colombi S., Juszkiewicz R., 1995, A \& A, 298, 643-660
\bibitem[\protect\citeauthoryear{Hivon, G\'orski, Netterfield, Brill, Prunet and Hansen}{2002}]{Hivon02} Hivon E., G\'orski K. M., Netterfield C. B., Brill B. P., Prunet S., Hansen F., 2002, APJ, 567, 2, arXiv:astro-ph/0105302
\bibitem[\protect\citeauthoryear{Huchra et al}{2011}]{Huchra11} Huchra J. P., et al., 2011, arXiv:1108.0669 
\bibitem[\protect\citeauthoryear{Jackson}{1972}]{Jackson72} Jackson J. C., 1972, MNRAS, 156, 1P
\bibitem[\protect\citeauthoryear{Jain and Bertschinger}{1994}]{Jain94} Jain B., Bertschinger E., 1994, ApJ, 431, 495-505, arXiv:astro-ph/9311070
\bibitem[\protect\citeauthoryear{Jeong and Komatsu}{2006}]{Jeong06} Jeong D., Komatsu E., 2006, ApJ, 651, 619-626, arXiv:astro-ph/0604075
\bibitem[\protect\citeauthoryear{Jeong and Komatsu}{2009}]{Jeong09} Jeong, D., Komatsu, E., 2009, ApJ, 691, 569, arXiv:0805.2632
\bibitem[\protect\citeauthoryear{Juszkiewicz, Hellwing and van de Weygaert}{2012}]{Juszkiewicz12} Juszkiewicz R., Hellwing W.A., van de Weygaert R., 2012, arXiv:1205.6163v1 
\bibitem[\protect\citeauthoryear{Kaiser}{1987}]{Kaiser87} Kaiser N., 1987, MNRAS, 227, 1
\bibitem[\protect\citeauthoryear{Komatsu et al}{2011}]{WMAP7} Komatsu E., et al., ApJS, 192, 18, arXiv:1001.4538
\bibitem[\protect\citeauthoryear{Lanusse, Rassat and Starck}{2012}]{Lanusse12} Lanusse F., Rassat A., Starck J.-L., 2012, A\&A, 540, A92, arXiv:1112.0561
\bibitem[\protect\citeauthoryear{Laureijs et al}{2011}]{Laureijs11} Laureijs R., et al., 2011, arXiv:1110.3193 
\bibitem[\protect\citeauthoryear{Lazkoz, Maartens and Majerotto}{2006}]{Lazkoz06} Lazkoz R., Maartens R., Majerotto E., 2006, PRD, 74, 083510, arXiv:astro-ph/0605701
\bibitem[\protect\citeauthoryear{Leistedt et al}{2011}]{Leistedt11} Leistedt B., Rassat A., Refregier A., Strack J. -L., 2011, arXiv:1111.3591 
\bibitem[\protect\citeauthoryear{Linder}{2005}]{Linder05} Linder E. V., 2005, PRD, 72, 043529
\bibitem[\protect\citeauthoryear{Makino, Sasaki and Suto}{1992}]{Makino92} Makino N., Sasaki M., Suto Y., 1992, PRD, 68, 46, 585
\bibitem[\protect\citeauthoryear{Matsubara}{2008a}]{Matsubara08a} Matsubara T., 2008, PRD, 77, 063530
\bibitem[\protect\citeauthoryear{Matsubara}{2008b}]{Matsubara08b} Matsubara T., 2008, PRD, 78, 083519
\bibitem[\protect\citeauthoryear{Matsubara}{2011}]{Matsubara11} Matsubara T., 2011, PRD, 83, 083518
\bibitem[\protect\citeauthoryear{Nishimichi et al}{2007}]{Nishimichi07} Nishimichi T., et al., 2007, PASJ, 59, 1049, arXiv:0705.1589
\bibitem[\protect\citeauthoryear{Nishimichi et al}{2009}]{Nishimichi09} Nishimichi T., et al., 2009, Publ. Astron. Soc. Jpn., 61, 321, arXiv:0810.0813
\bibitem[\protect\citeauthoryear{Nock, Percival and Ross}{2010}]{Nock10} Nock K., Percival W. J., Ross A. J., 2010, MNRAS, 407, 520, arXiv:1003.0896
\bibitem[\protect\citeauthoryear{Nomura, Yamamoto and Nishimichi}{2008}]{Nomura08} Nomura H., Yamamoto K., Nishimichi T., 2008, JCAP, 0810, 031, arXiv:0809.4538
\bibitem[\protect\citeauthoryear{Nomura, Yamamoto, Huetsi and Nishimichi}{2009}]{Nomura09} Nomura H., Yamamoto K., Huetsi G., Nishimichi T., 2009, PRD, 79, 063512, arXiv:0903.1883
\bibitem[\protect\citeauthoryear{Okumura and Jing}{2011}]{Okumura11} Okumura T., Jing Y. P., 2011, APJ, 726, 5
\bibitem[\protect\citeauthoryear{Okamura, Taruya and Matsubara}{2011}]{Okamura11b} Okamura T., Taruya A., Matsubara T., 2011, JCAP, 1108, 012
\bibitem[\protect\citeauthoryear{Padmanabhan et al}{2005}]{Padmanabhan05} Padmanabhan N., et al., 2005, MNRAS, 359, 237, arXiv:astro-ph/0407594 
\bibitem[\protect\citeauthoryear{Padmanabhan, White and Cohn}{2009}]{Padmanabhan09} Padmanabhan N., White M., Cohn J. D., 2009, PRD, 79, 063523, arXiv:0812.2905
\bibitem[\protect\citeauthoryear{Peebles and Yu}{1970}]{Peebles70} Peebles P. J. E., Yu J. T., 1970, APJ, 162, 815
\bibitem[\protect\citeauthoryear{Percival et al}{2004}]{Percival04} Percival W. J., Burkey D., Heavens A., et al., 2004, MNRAS, 353, 1201, arXiv:astro-ph/0406513 
\bibitem[\protect\citeauthoryear{Percival et al}{2007}]{Percival07} Percival W. J., Cole S., Eisenstein D. J., Nichol R.C., Peacock J. A., Pope A. C., Szalay A. S., 2007, MNRAS, 381, 1053, arXiv:0705.3323
\bibitem[\protect\citeauthoryear{Percival et al}{2007}]{Percival07b} Percival W. J., et al., 2007, ApJ, 657, 51, arXiv:astro-ph/0608635
\bibitem[\protect\citeauthoryear{Rassat et al}{2008}]{Rassat08} Rassat A., et al., 2008, arXiv:0810.0003v1 [astro-ph]
\bibitem[\protect\citeauthoryear{Rassat and Refregier}{2012}]{Rassat12} Rassat A., Refregier A., 2012, arXiv:1112.3100 
\bibitem[\protect\citeauthoryear{Ross, Percival, Crocce, Cabré and Gaztanaga}{2011}]{Ross11} Ross A. J., Percival W. J., Crocce M., Cabré A., Gaztanaga E., 2011, MNRAS, 415, 3, 2193-2204, arXiv:1102.0968
\bibitem[\protect\citeauthoryear{Samushia et al}{2011}]{Samushia11} Samushia L., et al., 2011, MNRAS, 410, 1993-2002
\bibitem[\protect\citeauthoryear{Sato and Matsubara}{2011}]{Sato11} Sato M., Matsubara T., 2011, PRD, 84, 043501
\bibitem[\protect\citeauthoryear{Seo and Eisenstein}{2003}]{Seo03} Seo H. -J., Eisenstein D. J., 2003, ApJ, 598, 720, arXiv:astro-ph/0307460
\bibitem[\protect\citeauthoryear{Seo and Eisenstein}{2007}]{Seo07} Seo H. -J., Eisenstein D. J., 2007, ApJ, 665, 14, arXiv:astro-ph/0701079
\bibitem[\protect\citeauthoryear{Scoccimarro and Frieman}{1996}]{Scoccimarro96} Scoccimarro R., Frieman J., 1996, ApJ, 473, 620, astro-ph > arXiv:astro-ph/9602070
\bibitem[\protect\citeauthoryear{Scoccimarro, Couchman and Frieman}{1999}]{Scoccimarro99} Scoccimarro R., Couchman H. M. P., Frieman J. A., 1999, ApJ, 517, 531
\bibitem[\protect\citeauthoryear{Scoccimarro}{2004}]{Scoccimarro04} Scoccimarro R., 2004, PRD, 70, 083007 
\bibitem[\protect\citeauthoryear{Shapiro, Crittenden and Percival}{2011}]{Shapiro11} Shapiro C., Crittenden R. G., Percival W. J., 2011, MNRAS, 422, 2341-2350, arXiv:1109.1981
\bibitem[\protect\citeauthoryear{Shaw and Lewis}{2008}]{Shaw08} Shaw J. R., Lewis A., 2008, PRD, 78, 103512, arXiv:0808.1724
\bibitem[\protect\citeauthoryear{Smith, Scoccimarro and Sheth}{2008}]{Smith08} Smith R. E., Scoccimarro R., Sheth R. K., 2008, PRD, 77, 043525, arXiv:astro-ph/0703620
\bibitem[\protect\citeauthoryear{Slosar, Ho, White and Louis}{2009}]{Slosar09} Slosar A., Ho S., White M., Louis T., 2009, JCAP, 10, 19, arXiv:0906.2414
\bibitem[\protect\citeauthoryear{Sunyaev and Zeldovich}{1970}]{Sunyaev70} Sunyaev R. A., Zeldovich Y. B., 1970, apss, 7, 3
\bibitem[\protect\citeauthoryear{Suto and Sasaki}{1991}]{Suto91} Suto Y., Sasaki M., 1991, PRL, 66, 264
\bibitem[\protect\citeauthoryear{Taruya, Nishimichi, Saito and Hiramatsu}{2009}]{Taruya09} Taruya A., Nishimichi T., Saito S., Hiramatsu T., 2009, PRD, 80, 123503, arXiv:0906.0507
\bibitem[\protect\citeauthoryear{Taruya, Nishimichi and Saito}{2010}]{Taruya10} Taruya A., Nishimichi T., Saito S., 2010, PRD, 82, 063522, arXiv:1006.0699
\bibitem[\protect\citeauthoryear{de la Torre and Guzzo}{2012}]{Torre12} de la Torre S., Guzzo L., 2012, MNRAS, 427, 327-342
\bibitem[\protect\citeauthoryear{Umeh, Clarkson and Maartens}{2012}]{Umeh12} Umeh O., Clarkson C., Maartens R., 2012, arXiv:1207.2109, arXiv:1207.2109
\bibitem[\protect\citeauthoryear{Vishniac}{1983}]{Vishniac83} Vishniac E., 1983, MNRAS, 203, 345
\bibitem[\protect\citeauthoryear{Wang and Mukherjee}{2006}]{Wang06} Wang Y., Mukhurjee P., 2006, APJ, 650, 1, arXiv:astro-ph/0604051
\bibitem[\protect\citeauthoryear{Wang and Steinhardt}{1998}]{Wang98} Wang L., Steinhardt P. J., 1998, APJ, 508, 483
\bibitem[\protect\citeauthoryear{Xu et al}{2010}]{Xu10} Xu X., et al., 2010, ApJ, 718, 1224, arXiv:1001.2324
\bibitem[\protect\citeauthoryear{York et al}{2000}]{York00} York D. G., et al., 2000, AJ, 120, 1579, arXiv:astro-ph/0006396
\bibitem[\protect\citeauthoryear{Yoo and Desjacques}{2013}]{Yoo13} Yoo J., Desjacques V., 2013, arXiv:1301.4501
\bibitem[\protect\citeauthoryear{Zaldarriaga, Seljak and Bertschinger}{1998}]{Zaldarriaga98} Zaldarriaga M., Seljak U., Bertschinger E., 1998, APJ, 494, 491
\bibitem[\protect\citeauthoryear{Zaroubi and Hoffman}{1996}]{Zaroubi96} Zaroubi S., Hoffman Y., 1996, APJ, 462, 25
\end{thebibliography}

\appendix

\section{Spherical Bessel Function}
\label{appendixA}
In this section we quickly outline some of the more useful properties of the spherical Bessel functions 
that have been used in the derivation of our results. 
The first important property of spherical Bessel functions is that they obey a well-known orthogonality condition:

\be
\int_0^{\infty} r^2dr j_{\ell} (kr)j_{\ell}(k'r) = {\pi \over 2k k'}\delta(k-k') .
\ee

\n
The first derivative of the spherical Bessel function can be expressed 
using the following recursion relation:
\be
j_{\ell}'(r)= {1 \over 2\ell +1}\Big [\ell j_{\ell-1}(r) - (\ell +1)j_{\ell +1}(r)\Big ].
\label{eq:bess1}
\ee
The second- and higher-order derivatives are deduced by successive application
of the above expression: 
\ben
&&j_{\ell}''(r)= \Big [{(2l^2 +2 \ell -1) \over (2\ell +3)(2\ell +1)}j_{\ell}(r) - \nn \\
&& -{\ell (\ell -1) \over (2\ell -1)(2\ell +1) }j_{\ell -2}(r) - {(\ell +1)(\ell +2)\over (2\ell +1)(2\ell +3)}j_{\ell +2}(r)   \Big ].
\label{eq:bess2}
\een
These expressions can be used to simply the kernels $I_{\ell}^{(1)}(k,k')$ defined in 
Eqn.(\ref{eqn:Il1}) to express mode-mixing due to redshift-space distortion.

\section{Spherical Harmonics}
\label{appendixB}
The spherical-harmonics are complete and orthogonal on the surface of the sphere:
\ben
&& \sum_{\ell m} Y_{\ell m}(\oh) Y_{\ell m}(\oh') = \delta^{2\rm D}(\oh-\oh') \, ; \\ 
&&\int\; d\oh\; Y_{\ell m}(\oh)Y_{\ell' m'}(\oh') = \delta^{\rm K}_{\ell \ell'}\delta^{\rm K}_{mm'}.
\een
The overlap integrals of three spherical harmonics are given by the Gaunt integral which
are expressed in terms of 3j symbols (denoted by matrices below):
\ben
&&\int d\oh Y_{\ell m}(\oh)Y_{\ell' m'}(\oh)Y_{\ell'' m''}(\oh) = I_{\ell_1 \ell_2 \ell_3}\nn  \\
&&\quad\quad \times \left ( \begin{array}{ c c c }
     \ell_1 & \ell_2 & \ell_3 \\
     0 & 0 & 0
  \end{array} \right)
\left ( \begin{array}{ c c c }
     \ell_1 & \ell_2 & \ell_3 \\
     m_1 & m_2 & m_3
  \end{array} \right) \, ; \\
&& I_{\ell_1 \ell_2 \ell_3} = \sqrt{(2\ell_1 +1)(2\ell_2 +1)(2\ell_3 +1)\over 4\pi}.
\label{eq:Gaunt}
\een

\section{3J Symbols}
The following orthogonality properties of $3j$ symbols were used to simplify various expressions:
\ben
\sum_{l_3m_3} (2l_3+1) \left ( \begin{array}{ c c c }
     \ell_1 & \ell_2 & \ell_3 \\
     m_1 & m_2 & m_3
  \end{array} \right )
\left ( \begin{array}{ c c c }
     \ell_1 & \ell_2 & \ell \\
     m_1' & m_2' & m
  \end{array} \right )\nn \\
 = \delta^{\rm K}_{m_1 m_1'} \delta^{\rm K}_{m_2m_2'};
\een
\ben
 \sum_{m_1m_2} \left ( \begin{array}{ c c c }
     \ell_1 & \ell_2 & \ell_3 \\
     m_1 & m_2 & m_3
  \end{array} \right)
\left ( \begin{array}{ c c c }
     \ell_1 & \ell_2 & \ell_3' \\
     m_1 & m_2 & m_3'
  \end{array} \right)\nn \\
  = {\mathcal \delta^{\rm K}_{l_3l_3'} 
\delta^{\rm K}_{m_3m_3'} \over 2\ell_3 + 1} . 
\label{3j_ortho1}
\een

\section{Finite Surveys and Discrete Spherical Bessel-Fourier Transformation and Pseudo-$\myC_l$s}
\label{sec:pcl}
\subsection{3D Scalar fields}
Different types of boundary conditions are employed in the literature 
for finite surveys \citep{Binney91,Fisher95,Heavens95}.

A natural choice for the boundary condition is to assume that the field vanishes at the boundary
of the survey $r=R$ leading to following condition on the radial modes that is
determined by the zeros of the spherical Bessel functions $j_{\ell}(r)$:

\be
j_{\ell} (q_{\ell n})=j_{\ell}(k_{\ell n}R) = 0;\quad\quad q_{\ell n}=k_{\ell n}R.
\ee

The closure relation for spherical harmonics will take the following form:

\be
\int_0^1 \; dz\; z^2 j_{\ell} (k_{\ell n}z)j_{\ell} (k_{\ell n}z) 
= {1 \over 2}[j_{\ell+1}(q_{n \ell})]^2 \delta_{\ell \ell'}\delta_{nn'}. 
\ee

Which, in terms of the radial wavenumber, can be expressed as follows:

\be
\int_0^R\;dr\;r^2 k_{n \ell}k_{\ell' n'}j_l(k_{\ell n}r)j_{\ell'}(k_{\ell' n'}r) 
 ={k^2_{\ell n}[j_{\ell+1}(q_{\ell n})]^2 \over 2R^{-3}} \delta_{\ell \ell'}\delta_{nn'}.
\ee

The discrete spectrum is determined by the zeros of the spherical Bessel function. 
The normalisation coefficients are given by:
\be
{1 \over \tau_{n \ell}} = {R^3 \over 2}[ k_{n \ell}j_{\ell +1}(k_{\ell n}R)]^2.
\ee
The inverse and forward discrete sFB transforms are as follows:
\ben
&& \Psi_{\ell m}(k_{\ell n}) = \tau_{\ell n}\int d^3\br\; \Psi(\br) \, k_{\ell n} \, j_{\ell}(kr) \, Y_{\ell m}(\oh);\\
&& \Psi(\br) = \sum_{\ell mn} \tau_{\ell n}\Psi_{\ell m}(k)j_{\ell}(kr) Y_{\ell m}(\oh).
\een
The following expression is useful:
\ben
&& \Psi_{\ell m}(k_{\ell n}) = {i^{\ell } k_{\ell n} \over (2\pi)^{3/2}}\int d\oh_k \Psi(k_{\ell n}, \oh_k) Y_{\ell m}(\oh_k)
\een
In case of finite survey the 3D power-spectrum samples only discrete radial wave-numbers $k_{ln}$ 
which is defined by the survey radius $R$:
\ben
&& \la \Psi_{\ell m}(k_{\ell n})\Psi^*_{\ell' m'}(k_{\ell' n'}\ra = P_{\Psi\Psi}(k_{\ell n})\delta_{\ell \ell'}\delta_{mm'}\delta_{nn'} .
\een
In addition to finite survey size, surveys often have a mask $s(\oh)$. The sFB transform of
a masked field defines the convolved or {\em Pseudo} harmonics $\tilde\Psi_{lm}(k_{ln})$:
\ben
\tilde\Psi_{\ell m}(k_{\ell n}) = && \sqrt{2 \over \pi} \tau_{\ell n}\int_0^R r^2 dr \int_{\Omega} d\oh \nn \\
&& \times [\phi(r) s(\oh)] \Psi(\br) j_{\ell}(k_{\ell n}r) Y_{\ell m}(\oh)d\oh .
\een
The convolved or {\em Pseudo}-harmonics are expressed in terms of all-sky harmonics $\Psi_{lm}(k_{ln})$
by the following expression:
\ben
\tilde\Psi_{\ell m}(k_{\ell n}) =&& \sum_{n'} \sum_{\ell' m'}\sum_{\ell'' m''} \tau_{\ell' n'} W(k_{\ell n},k_{\ell' n'})\Psi_{\ell m}(k_{\ell' n'})\nn\\
&&\quad\quad\times s_{\ell'' m''} I_{\ell \ell' \ell''}\left ( \begin{array}{ c c c }
     \ell & \ell' & \ell'' \\
     m & m' & m''
  \end{array} \right) .
\label{eq:pseudo2}
\een
The kernel $W(k_{\ell n},k_{\ell' n'})$ depends on selection function $\phi(r)$:
\ben
&& W(k_{\ell n},k_{\ell' n'}) = \int_0^R \; r^2 \; dr\; \phi(r) j_{\ell}(k_{\ell n}r)j_{\ell}(k'_{\ell' n'}r) 
\een
The Pseudo-$\myC_{\ell}$s (PCLs) constructed from the convolved harmonics are a function of 
power spectrum of the angular mask $\myC_{\ell''}^\chi$, normalisation coefficients  $\tau_{\ell n}$
and the selection function $\phi$:
\ben
\tilde \myC_{\ell}(k_{\ell n}) && = \la\tilde\Psi_{\ell m}(k_{\ell n})\tilde\Psi^*_{\ell m}(k_{\ell n})\ra  \nn \\
&& = \sum_{n'} \sum_{\ell'}\sum_{\ell''} \tau^2_{\ell' n'} {I^2_{\ell \ell' \ell''} \over 2\ell +1}
\left ( \begin{array}{ c c c }
     \ell & \ell' & \ell'' \\
     0 & 0 & 0
  \end{array} \right)^2 \nn \\
&& \times W^2(k_{\ell n},k_{\ell' n'}) \myC_{\ell'}(k_{\ell' n'})\myC_{\ell''}^\chi .
\label{eq:pcl2}
\een
Notice that the PCLs $\tilde \myC_{\ell}(k_{\ell n})$ are linear superposition of the
power spectrum of underlying field $\myC_{\ell}(k_{\ell n})$. The mixing matrix $M_{\ell n,\ell 'n'}$ is given by:
\be
\tilde \myC_{\ell} (k_{\ell n}) = \sum_{\ell' n'}M_{\ell n,\ell' n'}\myC_{\ell}(k_{\ell' n'});
\ee
where the mixing matrix is given by the following expression:
\be
M_{\ell n,\ell' n'} = \sum_{\ell''} \tau^2_{\ell'n'} {I^2_{\ell \ell' \ell''} \over 2\ell+1}
\left ( \begin{array}{ c c c }
     \ell & \ell' & \ell'' \\
     0 & 0 & 0
  \end{array} \right)^2  \\
 W^2(k_{\ell n},k_{\ell' n'})\myC_{\ell''}^\chi.
\ee
An unbiased estimates of the 3D power spectra can be written as:
\be
\myC_{\ell}(k_{\ell n}) = \sum_{\ell' n'}M^{-1}_{\ell n,\ell' n'}\tilde\myC_{\ell}(k_{\ell' n'}).
\ee
This is an extension of well known results for the projected surveys \citep{Hivon02}.
For low sky-coverage and small survey volumes the matrix  $M_{\ell n, \ell' n'}$
is expected to be singular and binning of modes may be required.

A different choice of boundary condition is often employed \citep{Fisher95}:
\be
j_{\ell-1}(k'_{\ell n}R) =0;
\ee
The normalisation constants in this case are given by:
\be
{1 \over \tau_{\ell n}} = {R^3 \over 2}[ k_{\ell n}j_{\ell}(k_{\ell n}R)]^2.
\ee 
The expressions for the mixing matrix derived above can still
be used simply replacing the  normalisation coefficients $\tau_{\ell n}$.

For discrete fields such as the galaxy distribution 
we can use the PCL approach if we replace the continuous 
function $\Psi(\br)$ with a sum of delta functions that peak
at galaxy positions $r_s$: $\Psi(\br) = \sum_{s=1}^{\rm N}\delta^{3\rm D}(\br - \br_s)$;
here $N$ is the number of galaxies. The sFB for such a discrete field
is given by $\Psi_{\ell m}(k) = \sum_{s=1}^N \tau_{\ell m}j_{l}(r_sk_{\ell n})Y_{\ell m}(\oh_s)$.
Where the radial and angular position of galaxies are denoted by 
$\br_s = (r_s,\oh_s)= (r_s,\theta_s,\phi_s)$

\section{Standard Perturbation Theory}
\label{appendixSPT}
In the formalism outlined in section \ref{sec:pert}, any statistical observable can be computed to arbitrary order. Typically we are only interested in the 
second order corrections to the matter power spectrum though expressions for higher-order corrections have been derived. One of the key issues regarding the 
inclusion is the computational costs required for these higher-order corrections in part due to the high dimensionality of the integrals, even after symmetry 
arguments have been taken into account. The analytic expressions for the first corrections can be analytically derived \cite{Makino92}:

\begin{align}
 P_{13} (k) &= \frac{1}{252} \frac{k^3}{4 \pi^2} \int^{\infty}_0 dx P_{\textrm{lin}} (k) P_{\textrm{lin}} (k x) \Bigg[ \frac{12}{x^2} - 158 + 100 x^2 \nonumber  \\
 &\; \; - 42 x^4 + \frac{3}{x^2} \left(x^2 -1 \right)^3 \left( 7 x^2 + 2 \right) \log \left| \frac{1+x}{1 - x} \right| \Bigg] \\
 P_{22} (k) &= \frac{1}{98} \frac{k^3}{4 \pi^2} \int^{\infty}_0 dx P_{\textrm{lin}} (k x) \int^{1}_{-1} d \mu P_{\textrm{lin}} (k \sqrt{1 + x^2 - 2 x \mu}) \nonumber  \\
 &\; \; \times \frac{\left( 3 x + 7 \mu - 10 x \mu^2 \right)^2}{\left( 1 + x^2 - 2 x \mu \right)^2}  \\
 P^{\textrm{LPT}}_{13} (k) &= \frac{1}{252} \frac{k^3}{4 \pi^2} P_{\textrm{lin}} (k) \int^{\infty}_0 dx P_{\textrm{lin}} (kx) 
 \Bigg[ \frac{12}{x^2} + 10 + 100 x^2 \nonumber  \\
  &\; - 42 x^4 + \frac{3}{x^3} ( x^2 - 1)^3 (7 x^2 + 2) \log \Bigg| \frac{1 + x}{1 - x} \Bigg| \Bigg] .
\end{align}

\section{Flat Sky Limit}
For surveys that cover large opening angles on the sky, the full sFB expansion detailed above is the most natural and convenient choice. This 
expansion does, however, break down for small-angle surveys where the signal of interest occurs at high-$\ell$ modes. In such a situation the accurate computation of 
high-$\ell$ spherical harmonics is cumbersome and computationally expensive. Instead it is much more natural to approximate the spherical harmonics as sums of 
exponentials corresponding to a 2D Fourier expansion. Essentially we are replacing the spherical harmonics solutions with a plane-wave approximation valid at 
high multipoles. 

In the flat sky limit we expand a 3D field $\Psi$ at a 3D position ${\bf{r}} \equiv ( r , \vec{\theta} \,)$ on the sky using a basis consisting of 2D 
Fourier modes and radial Bessel functions:

\begin{align}
 f( r , \vec{\theta} \, ) &= \sqrt{\frac{2}{\pi}} \int k dk \int \frac{ d^2 \vec{\ell}}{(2 \pi)^2} \, f(k,\vec{\ell} \, ) \, j_{\ell} (kr) \, e^{i 
 \vec{\ell} \cdot \vec{\theta}} \label{eqn:FlatSkyExp}\\
 f(k, \vec{\ell} \, ) &= \sqrt{\frac{2}{\pi}} \int r^2 dr \int d^2 \theta \, f(r,\vec{\theta} \, ) \, k \, j_{\ell} (kr) \, e^{-i \vec{\ell} \cdot \vec{\theta}}
\end{align}

\n
where ${\bf{\ell}}$ is a 2D angular wavenumber and k is a conventional radial wavenumber. We can simplify the analysis by adopting coordinates such that the 
survey corresponds to small angles around the pole of the spherical coordinates, defined by angles $(\theta , \phi)$ for which, in the limit $\theta \rightarrow 0$
, we can apply a 2D expansion of the plane waves:

\begin{equation}
 e^{i \vec{\ell} \cdot \vec{\theta}} \simeq \sqrt{\frac{2 \pi}{\ell}} \displaystyle\sum_m i^m Y_{\ell m} (\theta , \phi) e^{- i m \varphi_{\ell}} 
 \label{eqn:2DPlaneWaveExp}
\end{equation}

\n
where $\vec{\ell} = (\ell \cos \varphi_{\ell} , \ell \sin \varphi_{\ell})$ and $\vec{\theta} = (\theta \cos \varphi , \theta \sin \varphi)$. The correspondence 
between the 3D flat-sky and 3D full-sky coefficients can be obtained by substituting Eq.(\ref{eqn:2DPlaneWaveExp}) into Eq.(\ref{eqn:FlatSkyExp}) and noting 
that $\int d^2 \vec{\ell} = \int \ell d \ell \int d \varphi_{\ell} \rightarrow \sum_{\ell} \ell \int d \varphi_{\ell}$ in the high-${\ell}$ limit. The 
correspondence can be shown to be:

\begin{align}
 f_{\ell m} (k) &= \sqrt{\frac{\ell}{2 \pi}} i^m \int \frac{ d \varphi_{\ell} }{(2 \pi)} e^{- i m \varphi_{\ell}} f(k,\vec{\ell} \, ) \\
 f( k,\vec{\ell} \, ) &= \sqrt{\frac{2 \pi}{\ell}} \displaystyle\sum_m i^{-m} f_{\ell m} (k) e^{i m \varphi_{\ell} }
\end{align}

\n
We now extend this analysis to RSD by constructing harmonics of a field $\Psi ({\bf{r}})$ in the flat-sky limit when convolved with a selection function $\phi (s)$. 
These new flat-sky harmonics take into account the RSD much as before:

\begin{equation}
 \tilde{\Psi} (k,\vec{\ell} \, ) = \sqrt{\frac{2}{\pi}} \int s^2 ds \int d^2 \theta \, k \, \Psi (r, \vec{\theta} \, ) 
 \left[ \phi (s) j_{\ell} (ks) \right] e^{- i \vec{\ell} 
 \cdot \vec{\theta}} .
\end{equation}

\n
Following the same perturbative procedure results in a series expansion in $\beta$ where:

\begin{equation}
 \tilde{\Psi}_{\ell} (k , \vec{\ell} \, ) = \tilde{\Psi}^{(0)}_{\ell} (k , \vec{\ell} \, ) + \tilde{\Psi}^{(1)}_{\ell} (k , \vec{\ell} \, ) + \dots .
\end{equation}

\n
As before the $\tilde{\Psi}^{(0)}_{\ell} (k , \vec{\ell} \, )$ term represents the unredshifted contribution:

\begin{equation}
 \tilde{\Psi}^{(0)}_{\ell} (k , \vec{\ell} \, ) = \sqrt{\frac{2}{\pi}} \int r^2 dr \int d^2 \theta \, \Psi (r, \vec{\theta} \, ) 
 \, k \, \left[ j_{\ell} (kr) \phi (r) \right] 
 e^{-i \vec{l} \cdot \vec{\theta}} 
\end{equation}

\begin{align}
 \tilde{\Psi}^{(1)}_{\ell} (k , \vec{\ell} \, ) &= \sqrt{\frac{2}{\pi}} \int r^2 dr \int d^2 \theta \, \Psi (r, \vec{\theta} \, ) 
 \, k \nonumber  \\ 
 & \times \left\lbrace \left[ {\bf{v}}(\vec{r}) \cdot \vec{\theta} \right]  \frac{d}{dr} \left[ j_{\ell} (kr) \psi (r) \right] \right\rbrace 
 e^{-i \vec{\ell} \cdot \vec{\theta}} 
\end{align}

\end{document}